\documentclass[traditabstract]{aa}  

\usepackage{psfig}

\begin{document}

\title{BL Lacertae identifications in a {\it ROSAT}-selected sample of 
{\it Fermi} unidentified objects\thanks{Based on observations collected at 
the following telescopes: Telescopio Nazionale Galileo of the Observatorio 
del Roque de los Muchachos of the Instituto de Astrof\'{\i}sica de 
Canarias (Canary Islands, Spain); New Technology Telescope of the European 
Southern Observatory (ESO) in Cerro La Silla (Chile), under program 
89.A-0148(A); Unit 2 `Kueyen' of the Very Large Telescope of the 
ESO in Cerro Paranal (Chile), under program 71.D-0176(A); 1.5-meter 
telescope of the Cerro Tololo Interamerican Observatory (Chile).}}


\author{N. Masetti\inst{1},
B. Sbarufatti\inst{2},
P. Parisi\inst{3},
E. Jim\'enez-Bail\'on\inst{4},
V. Chavushyan\inst{5},
F.P.A. Vogt\inst{6},
V. Sguera\inst{1},
J.B. Stephen\inst{1},
E. Palazzi\inst{1},
L. Bassani\inst{1},
A. Bazzano\inst{3}, 
M. Fiocchi\inst{3},
G. Galaz\inst{7},
R. Landi\inst{1},
A. Malizia\inst{1},
D. Minniti\inst{7,8},
L. Morelli\inst{9,10} and
P. Ubertini\inst{3}
}

\institute{
INAF -- Istituto di Astrofisica Spaziale e Fisica Cosmica di 
Bologna, Via Gobetti 101, I-40129 Bologna, Italy
\and
Department of Astronomy and Astrophysics, Pennsylvania State University, 
University Park, PA 16802, USA
\and
INAF -- Istituto di Astrofisica e Planetologia Spaziali, Via Fosso del 
Cavaliere 100, I-00133 Rome, Italy
\and
Instituto de Astronom\'{\i}a, Universidad Nacional Aut\'onoma de M\'exico,
Apartado Postal 70-264, 04510 M\'exico D.F., M\'exico
\and
Instituto Nacional de Astrof\'{i}sica, \'Optica y Electr\'onica,
Apartado Postal 51-216, 72000 Puebla, M\'exico
\and
Research School of Astronomy and Astrophysics, Australian National 
University, Canberra ACT-2611, Australia
\and
Instituto de Astrof\'{i}sica, Pontificia Universidad Cat\'olica de Chile, 
Casilla 306, Santiago 22, Chile
\and
Specola Vaticana, V-00120 Citt\`a del Vaticano
\and
Dipartimento di Fisica ed Astronomia ``G. Galilei", Universit\`a di Padova,
Vicolo dell'Osservatorio 3, I-35122 Padua, Italy
\and
INAF-Osservatorio Astronomico di Padova, Vicolo dell'Osservatorio 5, 
I-35122 Padua, Italy
}

\offprints{N. Masetti (\texttt{masetti@iasfbo.inaf.it)}}
\date{Received 5 September 2013; accepted 23 September 2013}

\abstract{The optical spectroscopic followup of 27 sources belonging to a 
sample of 30 high-energy objects selected by positionally cross 
correlating the first {\it Fermi}/LAT Catalog and the {\it ROSAT} All-Sky 
Survey Bright Source Catalog is presented here. It has been found or 
confirmed 
that 25 of them are BL Lacertae objects (BL Lacs), while the remaining two 
are Galactic cataclysmic variables (CVs). This strongly suggests that the 
sources in the first group are responsible for the GeV emission 
detected with {\it Fermi}, while the two CVs most likely represent 
spurious associations. We thus find an 80\% {\it a posteriori} probability 
that the sources selected by matching GeV and X--ray catalogs belong to 
the BL Lac class. We also show suggestions that the BL Lacs 
selected with this approach are probably high-synchrotron-peaked sources 
and in turn good candidates for the detection of ultra-high-energy (TeV) 
photons from them.}


\titlerunning{BL Lacs in a {\it Fermi}/{\it ROSAT} sample}
\authorrunning{N. Masetti et al.}

\maketitle

\section{Introduction}

Until the end of the first decade of the present century the only 
available survey of the $\gamma$--ray sky was the one performed by EGRET 
(Hartman et al. 1999; Casandjian \& Grenier 2008). One of the main 
problems with this survey is that about half of the detected high-energy 
sources (170 out of 271 in Hartman et al. 1999, and 87 out of 188 in 
Casandjian \& Grenier 2008) could not be associated with longer-wavelength 
objects. This was mainly due to the large error boxes of these objects 
(with sizes of tens of arcminutes or more) which did not help in 
pinpointing their actual counterparts.

Therefore, one of the main objectives of the {\it Fermi} mission (McEnery 
et al. 2012), launched on June 2008 is the survey of the whole sky at 
$\gamma$--ray energies in the MeV-GeV range making use of the large 
collecting area and field of view of its instruments. In particular, the 
Large Area Telescope (LAT; Atwood et al. 2009) allows the emission of 
$\gamma$--ray objects to be detected in the 0.1--100 GeV band and permits 
their localization with an accuracy of a few arcminutes, depending on the 
significance and the spectral hardness with which the sources are detected.

After a preliminary issue of a list of bright $\gamma$--ray objects (Abdo 
et al. 2009), two catalogs of {\it Fermi}/LAT sources have been published 
by Abdo et al. (2010a) and Nolan et al. (2012) by accumulating 11 and 24 
months of LAT observations, respectively. In the first, 1451 objects are 
present, 821 of which are associated with known sources at other 
wavelengths (Abdo et al. 2010a). In the second LAT catalog, Nolan et al. 
(2012) reported 1873 $\gamma$--ray sources, with 1297 of them identified 
with certainty or probably associated with counterparts of $\gamma$--ray 
producing source classes.

From the analysis of the position and the $\gamma$--ray timing and 
spectral behavior of the sources in these two surveys it is found that 
most of them are associated with extragalactic objects including blazars, 
flat spectrum radio quasars and/or BL Lacertae objects (BL Lacs), or with 
Galactic sources such as pulsars, pulsar wind nebulae, supernova remnants 
(SNRs), globular clusters, and a few $\gamma$--ray binaries. Some peculiar 
objects are also found, but their association is less firm. No tentative 
or credible counterpart is, however, found for a number of these cataloged 
$\gamma$--ray sources, which amount to $\sim$30--40\% depending on the 
chosen catalog.

The identification of these objects is nevertheless crucial for several 
investigations. In particular, the detection of high--energy blazars is of 
paramount importance to determine, for instance, a more accurate 
measurement of the imprint of the extragalactic background light in their 
$\gamma$--ray spectra (e.g., Ackermann et al. 2012a) and to improve the 
estimate of the blazar contribution to the extragalactic $\gamma$--ray 
background (e.g., Abdo et al. 2010b). In addition, having information 
about these unidentified $\gamma$-ray sources can give tight constraints 
on dark matter models (e.g., Zechlin et al. 2012; Massaro et al. 2013a).

The main issue to tackle for the identification of the nature of these 
objects is, again, the relatively large error box size of LAT sources (a 
few arcminutes in radius at least). Several approaches to attack this 
problem on both statistical grounds (e.g., Munar-Adrover et al. 2011; 
Ackermann et al. 2012b; D'Abrusco et al. 2013) and observational grounds 
(e.g., Mirabal \& Halpern 2009; Shaw et al. 2012, 2013) were applied in 
recent years.

In particular, Stephen et al. (2010) tried a mixed 
statistical-observational approach by cross-correlating the positions of 
the first {\it Fermi}/LAT Catalog with those in the {\it ROSAT} All-Sky 
Survey Bright Source Catalog (Voges et al. 1999), which mapped the soft 
X--ray sky in the 0.1--2.4 keV range. The authors found a strong level of 
positional correlation between unassociated {\it Fermi}/LAT sources and 
{\it ROSAT} cataloged objects, leading to evidence for the association of 
a number of $\gamma$--ray sources with a soft X--ray counterpart, better 
positions (down to a few arcseconds) for all correlated objects, and hence 
the possibility of follow-up work at longer wavelengths, especially in 
optical.

In detail, Stephen et al. (2010) found 30 secure {\it Fermi}-{\it ROSAT} 
associations which allowed a likely optical counterpart to be pinpointed 
for nearly all of them. They also gave a tentative determination of the 
nature of these sources on the basis of their archival multiwavelength 
properties, finding that most associations (25 out of 30) are with blazar 
candidates, while only two are associated with Galactic objects (a SNR and 
a $\gamma$--ray binary with a low--mass companion star).

In order to complete this exploratory work which merges the $\gamma$--ray 
knowledge with the soft X--ray knowledge we decided to extend this 
information by performing an optical spectroscopy follow-up campaign on 
the targets with unidentified or tentatively proposed nature which were 
selected through the procedure of Stephen et al. (2010).

The present paper is structured as follows. In Sect. 2 we outline the 
sample of the selected sources and the additional techniques we used to 
refine their localization. In Sect. 3 a description of the observations is 
given. Section 4 presents the results, and a discussion about them is 
given in Sect. 5; the conclusions are shown in Sect. 6.

\section{The selected sample}

\begin{table*}[th!]
\caption[]{Log of the spectroscopic observations presented in this paper
(see text for details).}
\scriptsize
\begin{center}
\begin{tabular}{lllccr}
\noalign{\smallskip}
\hline
\hline
\noalign{\smallskip}
\multicolumn{1}{c}{{\it (1)}} & \multicolumn{1}{c}{{\it (2)}} & \multicolumn{1}{c}{{\it (3)}} & \multicolumn{1}{c}{{\it (4)}} & 
{\it (5)} & \multicolumn{1}{c}{{\it (6)}} \\
\multicolumn{1}{c}{Object} & \multicolumn{1}{c}{RA} & \multicolumn{1}{c}{Dec} & 
Telescope+instrument & UT Date \& Time & Exposure \\
 & \multicolumn{1}{c}{(J2000)} & \multicolumn{1}{c}{(J2000)} & & at mid-exposure & time (s) \\

\noalign{\smallskip}
\hline
\noalign{\smallskip}

1FGL J0051.4$-$6242 & 00:51:16.65     & $-$62:42:04.4     & NTT+EFOSC2  & 05 Sep 2012, 05:14 & 2$\times$1200 \\
1FGL J0054.9$-$2455 & 00:54:46.75     & $-$24:55:29.1     & TNG+DOLoReS & 17 Aug 2011, 03:21 & 2$\times$1200 \\
1FGL J0131.2+6121   & 01:31:07.23     &   +61:20:33.4     & TNG+DOLoReS & 05 Aug 2011, 02:50 & 2$\times$1800 \\
1FGL J0137.8+5814   & 01:37:50.47     &   +58:14:11.2     & TNG+DOLoReS & 14 Mar 2012, 21:06 & 2$\times$1200 \\
1FGL J0223.0$-$1118 & 02:23:14.26     & $-$11:17:38.8     & TNG+DOLoReS & 17 Aug 2011, 04:16 & 2$\times$1200 \\
1FGL J0506.9$-$5435 & 05:06:57.80     & $-$54:35:03.8     & NTT+EFOSC2  & 02 Oct 2012, 06:27 & 3$\times$1000 \\
1FGL J0604.2$-$4817 & 06:04:08.62     & $-$48:17:25.2     & NTT+EFOSC2  & 02 Oct 2012, 07:30 & 3$\times$1200 \\
1FGL J0648.8+1516   & 06:48:47.64     &   +15:16:24.8     & TNG+DOLoReS & 19 Feb 2011, 00:35 &  2$\times$900 \\
1FGL J0838.6$-$2828 & 08:38:43.37$^*$ & $-$28:27:01.5$^*$ & TNG+DOLoReS & 08 Feb 2011, 00:46 & 2$\times$1200 \\ 
1FGL J0841.4$-$3558 & 08:41:21.63     & $-$35:55:05.8     & NTT+EFOSC2  & 02 Oct 2012, 08:42 & 3$\times$1200 \\
1FGL J0848.6+0504   & 08:48:39.66$^*$ &   +05:06:17.9$^*$ & TNG+DOLoReS & 27 Feb 2011, 21:58 & 3$\times$1800 \\
1FGL J1304.3$-$4352 & 13:04:21.00     & $-$43:53:10.2     & NTT+EFOSC2  & 04 Sep 2012, 23:31 &          1200 \\
1FGL J1307.6$-$4259 & 13:07:37.97     & $-$42:59:38.9     & NTT+EFOSC2  & 04 Sep 2012, 23:59 &          1200 \\
1FGL J1419.7+7731   & 14:19:00.41     &   +77:32:29.6     & TNG+DOLoReS & 19 Feb 2011, 02:09 & 2$\times$1200 \\
1FGL J1544.5$-$1127 & 15:44:39.31$^*$ & $-$11:28:04.3$^*$ & TNG+DOLoReS & 04 Jun 2011, 02:53 & 2$\times$1800 \\ 
1FGL J1553.5$-$3116 & 15:53:33.54     & $-$31:18:31.4     & NTT+EFOSC2  & 05 Sep 2012, 00:41 & 2$\times$1200 \\
1FGL J1643.5$-$0646 & 16:43:28.91     & $-$06:46:19.5     & NTT+EFOSC2  & 05 Sep 2012, 02:14 & 2$\times$1200 \\
1FGL J1823.5$-$3454 & 18:23:38.59     & $-$34:54:12.0     & NTT+EFOSC2  & 05 Sep 2012, 03:06 & 2$\times$1200 \\
1FGL J1841.9+3220   & 18:41:47.05     &   +32:18:38.9     & TNG+DOLoReS & 23 Feb 2011, 05:58 &  2$\times$900 \\
1FGL J1926.8+6153   & 19:26:49.89     &   +61:54:42.3     & TNG+DOLoReS & 22 Mar 2011, 05:42 &  2$\times$900 \\
1FGL J1933.3+0723   & 19:33:20.30     &   +07:26:21.8     & TNG+DOLoReS & 02 Jun 2011, 03:54 &          1800 \\
1FGL J2042.2+2427   & 20:42:06.04     &   +24:26:52.3     & TNG+DOLoReS & 19 Ago 2011, 02:53 &  2$\times$900 \\
1FGL J2146.6$-$1345 & 21:46:36.94     & $-$13:44:00.6     & TNG+DOLoReS & 06 Jun 2011, 04:49 & 2$\times$1200 \\
1FGL J2323.0$-$4919 & 23:22:54.43     & $-$49:16:30.1     & NTT+EFOSC2  & 05 Sep 2012, 03:52 & 2$\times$1200 \\
1FGL J2329.2+3755   & 23:29:14.27     &   +37:54:14.6     & TNG+DOLoReS & 02 Jun 2011, 04:36 &  2$\times$900 \\

\noalign{\smallskip}
\hline
\noalign{\smallskip}
\multicolumn{6}{l}{Note: if not indicated otherwise, source coordinates were extracted from the 
2MASS catalog and have an accuracy better than 0$\farcs$1.}\\
\multicolumn{6}{l}{$^*$: coordinates extracted from the USNO catalogs, having 
an accuracy of about 0$\farcs$2 (Deutsch 1999; Assafin et al. 2001; Monet et al. 2003).}\\

\noalign{\smallskip}
\hline
\hline
\noalign{\smallskip}
\end{tabular}
\end{center}
\end{table*}

By considering the 30 objects listed in Table 1 of Stephen et al. (2010), 
we first performed a preliminary check to verify that the sources are 
recovered in the 2$^{\rm nd}$ {\it Fermi}/LAT Catalog (Nolan et al. 
2012); we determined that this is indeed true for all objects.

Next we took into account those sources in Stephen et al. (2010) with no 
identified nature or with identifications that had not yet been confirmed 
at the time of the publication of that work (October 2010). Thus, 1FGL 
J1910.9+0906 (associated with the SNR G043.3$-$00.2; Mezger et al. 1967) 
was not considered, nor was 1FGL J1227.9$-$4852 (the possible 
$\gamma$--ray counterpart of the atypical low-mass X--ray binary XSS 
J12270$-$4859; de Martino et al. 2010, 2013), along with the confirmed BL 
Lacs 1FGL J1353.6$-$6640 and 1FGL J1942.7+1033 (both observed with the ESO 
8-meter Unit 2 `Kueyen' of the Very Large Telescope by Tsarevsky et al. 
2005).

On the other hand, we kept in our selection the objects 1FGL J0131.2+6121 
and 1FGL J0137.8+5814 because, even if they were already spectroscopically 
identified as BL Lacs by Mart\'{i} et al. (2004) and Bikmaev et al. 
(2008), respectively, their published spectra had relatively poor 
signal-to-noise (S/N) ratios; so, we considered these sources as well in 
order to acquire spectroscopic data with higher quality. Thus 26 objects 
out of 30 in the original sample of Stephen et al. (2010) passed this 
check.

We then used the information gathered from radio catalogs such as those 
compiled with the NRAO VLA Sky Survey (NVSS; Condon et al. 1998), the 
Sydney University Molonglo Sky Survey (SUMSS; Mauch et al. 2003), and the 
Molonglo Galactic Plane Survey (MGPS; Murphy et al. 2007) to further 
reduce the {\it ROSAT} X--ray position uncertainty down to less than 2$''$ 
and so to determine the actual optical counterpart with a higher degree of 
accuracy. In parallel, to confirm the presence of X--ray emission from the 
position of the {\it ROSAT} sources, we browsed the {\it Swift}/XRT 
archive\footnote{XRT archival data are freely available at \\ {\tt 
http://www.asdc.asi.it/}} (see also Landi et al. 2010 for 1FGL 
J2056.7+4938/IGR J20569+4940) or, in the case of 1FGL J0137.8+5814, the 
{\it XMM-Newton} Serendipitous Source Catalog (Watson et al. 2009). In all 
cases we could confirm the soft X--ray emission detected with {\it ROSAT} 
and pinpoint a putative optical counterpart; moreover, for 24 objects, the 
additional presence of a positionally coincident radio source allowed the 
association of a single optical object beyond any reasonable doubt.

Concerning the above, we would like to spend a word about the case of 1FGL 
J0841.4$-$3558. Stephen et al. (2010) associated this source with the 
F-type bright star HD 74208 (=HIP 42640), with magnitude $V$ = 7.91 (H\o g 
et al. 2000). In addition to the {\it ROSAT} source, in the XRT data we 
found another soft X--ray object located about 1$\farcm$3 north of it: 
this object is detected above 3 keV and at radio frequencies, whereas the 
one associated with star HD 74208 is not. Although this strongly suggests 
that the former, harder X--ray emitter is the actual counterpart of 1FGL 
J0841.4$-$3558, for the sake of completeness we decided to perform 
spectroscopy on both optical sources associated with these two X--ray 
objects detected with {\it Swift}/XRT.

Unfortunately, the faintness ($i'$ = 19.5 mag; Mart\'{i} et al. 2012) of 
the optical counterpart of source 1FGL J2056.7+4938 as well as its 
closeness to the very bright star ($V$ = 8.82; H\o g et al. 2000) BD+49 
3420 prevented us from obtaining meaningful optical spectroscopy of it. In 
the end, we acquired optical spectroscopy on the selected putative 
counterparts of the sources listed in Table 1.

Given that all objects in Table 1 have $R$-band magnitudes mostly in the 
16$\div$18 range according to the U.S. Naval Observatory (USNO) catalogs 
(Monet et al. 2003; see also Table 2), we opted to perform optical 
spectroscopic observations with medium-sized (4-meter class) telescopes; 
only in the case of the bright star HD 74208 did we use a smaller, 
1.5-meter telescope (this observation is not reported in Table 1; see next 
section for details).

To complete the information about the {\it Fermi}/{\it ROSAT} sample of 
Stephen et al. (2010), we also retrieved from the ESO data 
archive\footnote{{\tt http://archive.eso.org/eso/eso\_archive\_main.html}} 
the relevant files concerning optical spectroscopy of the counterparts of 
sources 1FGL J1353.6$-$6640 and 1FGL J1942.7+1033 (also in this case, 
these two objects are not reported in Table 1; we refer the reader to 
Tsarevsky et al. 2005 for details about the corresponding observational 
setup) and reduced them independently. In total we analyzed the spectra of 
28 optical objects connected with 27 {\it Fermi}/{\it ROSAT} associations 
from Stephen et al. (2010).

\section{Optical observations and data analysis}

Optical spectroscopy of sources accessible from the northern hemisphere 
(i.e., with declination $\ga$ $-$30$^\circ$, with the single exception of 
1FGL J1643.5$-$0646) was acquired with the 3.58-meter Telescopio 
Nazionale Galileo (TNG) located at La Palma, Canary Islands (Spain). Its 
imaging spectrograph DOLoReS carried a 2048$\times$2048 pixel E2V 4240 
CCD; the spectra were acquired with the LR-B grism and a 1$\farcs$5-wide 
slit, which secured a nominal spectral coverage in the 3500--8200 
\AA~range and a dispersion of 2.5 \AA/pixel. The TNG data were acquired 
between February 2011 and March 2012.

We then acquired spectra of the southernmost objects (with declination 
below $-$30$^\circ$) plus 1FGL J1643.5$-$0646 again using a 3.58-meter 
telescope, the ESO New Technology Telescope (NTT) of the La Silla 
Observatory (Chile) equipped with the EFOSC2 instrument. Spectroscopic 
data were obtained with a 2048$\times$2048 pixel Loral/Lesser CCD. Grism 
\#13 and slits of width between 1$''$ and 2$''$ (depending on the night 
seeing) were used; this guaranteed a 3700--9300 \AA~nominal spectral 
coverage and a dispersion of 2.8 \AA/pixel. The data were acquired in 
September and October 2012.

Table 1 reports the log of these observations along with the coordinates 
of the observed optical sources, extracted from the Two Micron All Sky 
Survey (2MASS, with a positional uncertainty of $\leq$0$\farcs$1: 
Skrutskie et al. 2006) or the USNO catalogs (with accuracies of about 
0$\farcs$2: Deutsch 1999; Assafin et al. 2001; Monet et al. 2003).

As mentioned in the previous section, for the sake of completeness we also 
obtained spectroscopy of star HD 74208 with the 1.5-meter CTIO telescope 
of Cerro Tololo (Chile) equipped with the R-C spectrograph, which carries 
a 1274$\times$280 pixel Loral CCD. Two 60s spectroscopic frames were 
secured on 6 February 2012, with start times at 02:10 and 02:12 UT, 
respectively. Data were secured using Grating \#13/I and with a slit 
width of 1$\farcs$5, giving a nominal spectral coverage between 3300 and 
10500 \AA~and a dispersion of 5.7 \AA/pixel.

The whole set of spectroscopic data acquired at these telescopes, as well 
as the data from the ESO archive, was optimally extracted (Horne 1986) and 
reduced following standard procedures using IRAF\footnote{IRAF is the 
Image Reduction and Analysis Facility made available to the astronomical 
community by the National Optical Astronomy Observatories, which are 
operated by AURA, Inc., under contract with the U.S. National Science 
Foundation. It is available at {\tt http://iraf.noao.edu/}}.  Calibration 
frames (flat fields and bias) were taken on the day preceding or following 
the observing night. The wavelength calibration was performed using lamp 
data acquired soon after each on-target spectroscopic acquisiton; the 
uncertainty in this calibration was $\sim$0.5~\AA~in all cases according 
to our checks made using the positions of background night sky lines. Flux 
calibration was obtained using cataloged spectrophotometric standards. 
Finally, when multiple spectra were acquired from a given object, the data 
were stacked together to increase the S/N ratio.

Furthermore, given that all the considered objects except the putative 
counterpart of 1FGL J1823.5$-$3454 have an optical $R$-band magnitude 
tabulated in the USNO-A2.0\footnote{available at: \\ {\tt 
http://archive.eso.org/skycat/servers/usnoa}} catalog (see also Table 2), 
we used the $R$-band acquisition image corresponding to the spectroscopic 
observation of this source to obtain the relevant photometric information. 
The image was acquired with NTT plus EFOSC2 in imaging mode, which covers 
a field of 4$\farcm$1$\times$4$\farcm$1 with a scale of 0$\farcs$12 
pix$^{-1}$. The measurement was performed using simple aperture 
photometry, and calibrated using field stars with cataloged USNO-A2.0 
$R$-band magnitudes.

\section{Results}

\begin{figure*}
\mbox{\psfig{file=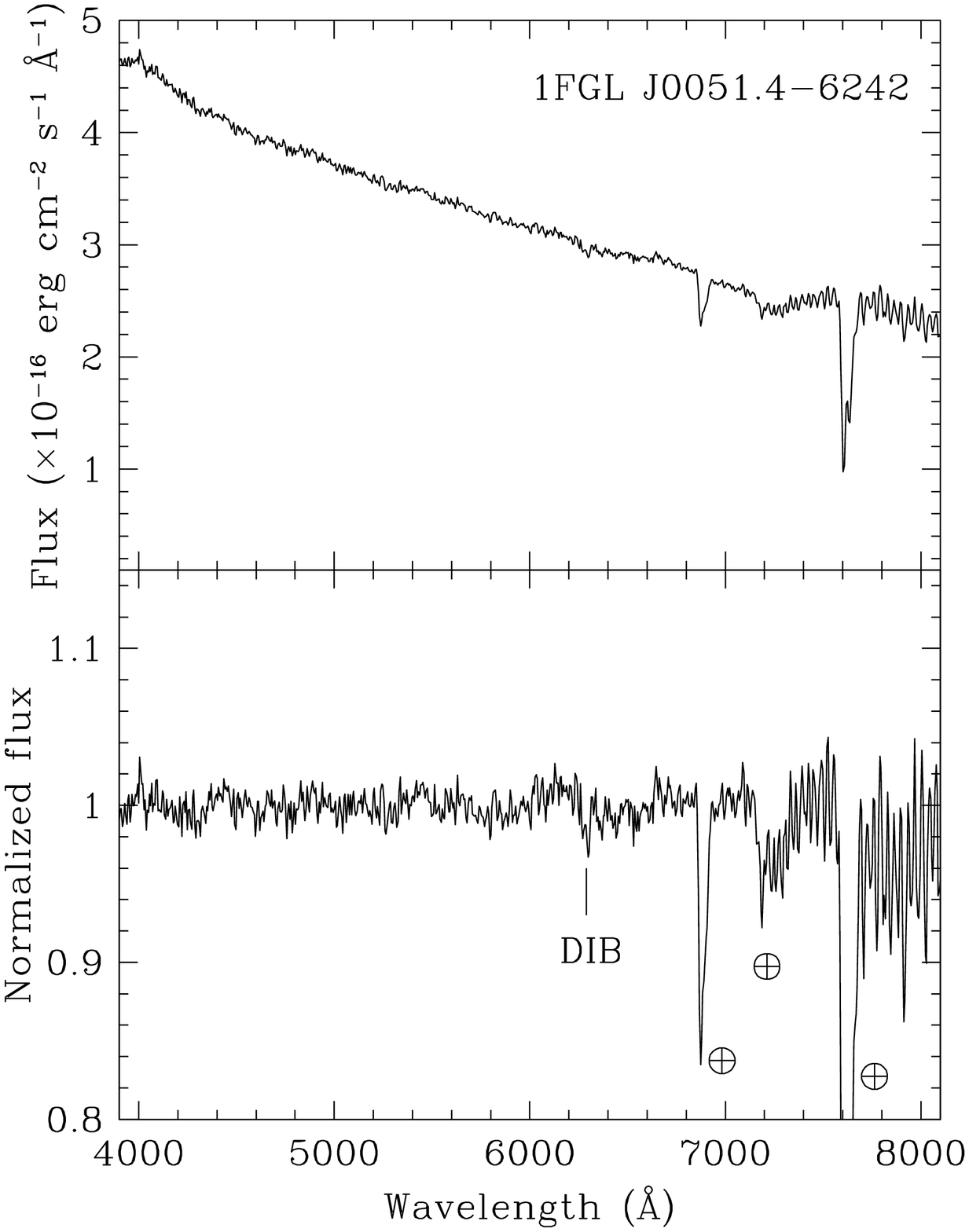,width=9cm,angle=0}}
\mbox{\psfig{file=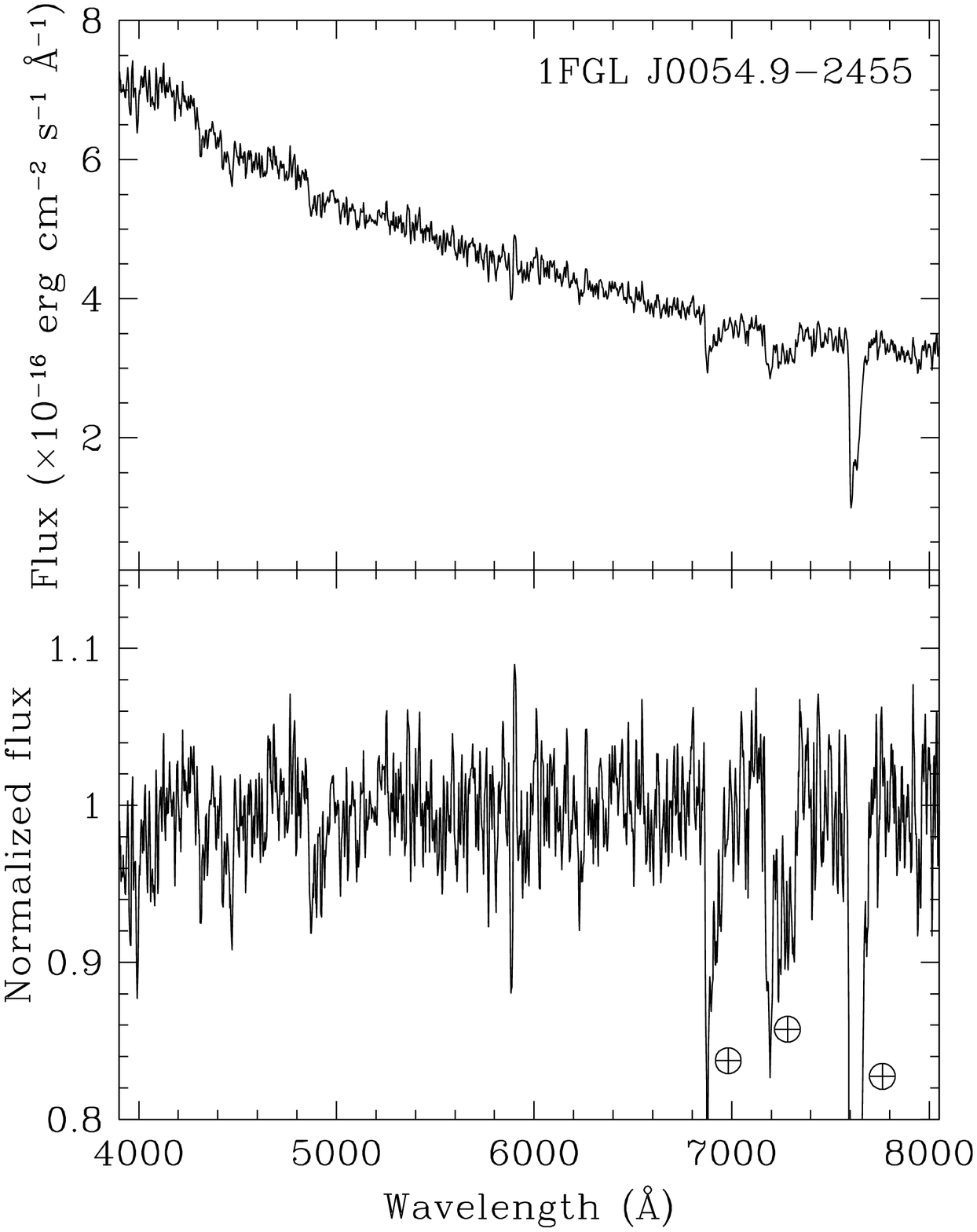,width=9cm,angle=0}}

\vspace{-.9cm}
\mbox{\psfig{file=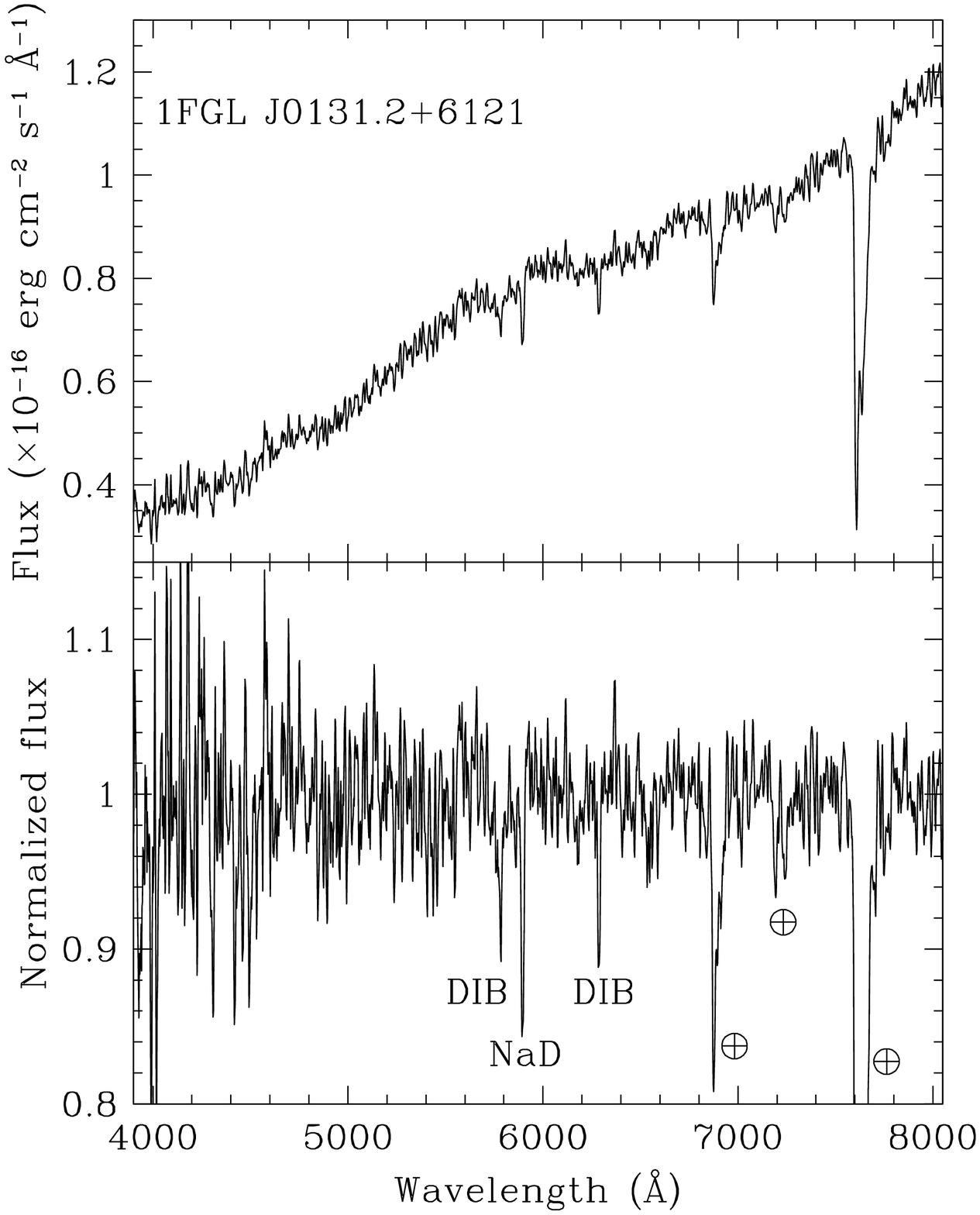,width=9cm,angle=0}}
\mbox{\psfig{file=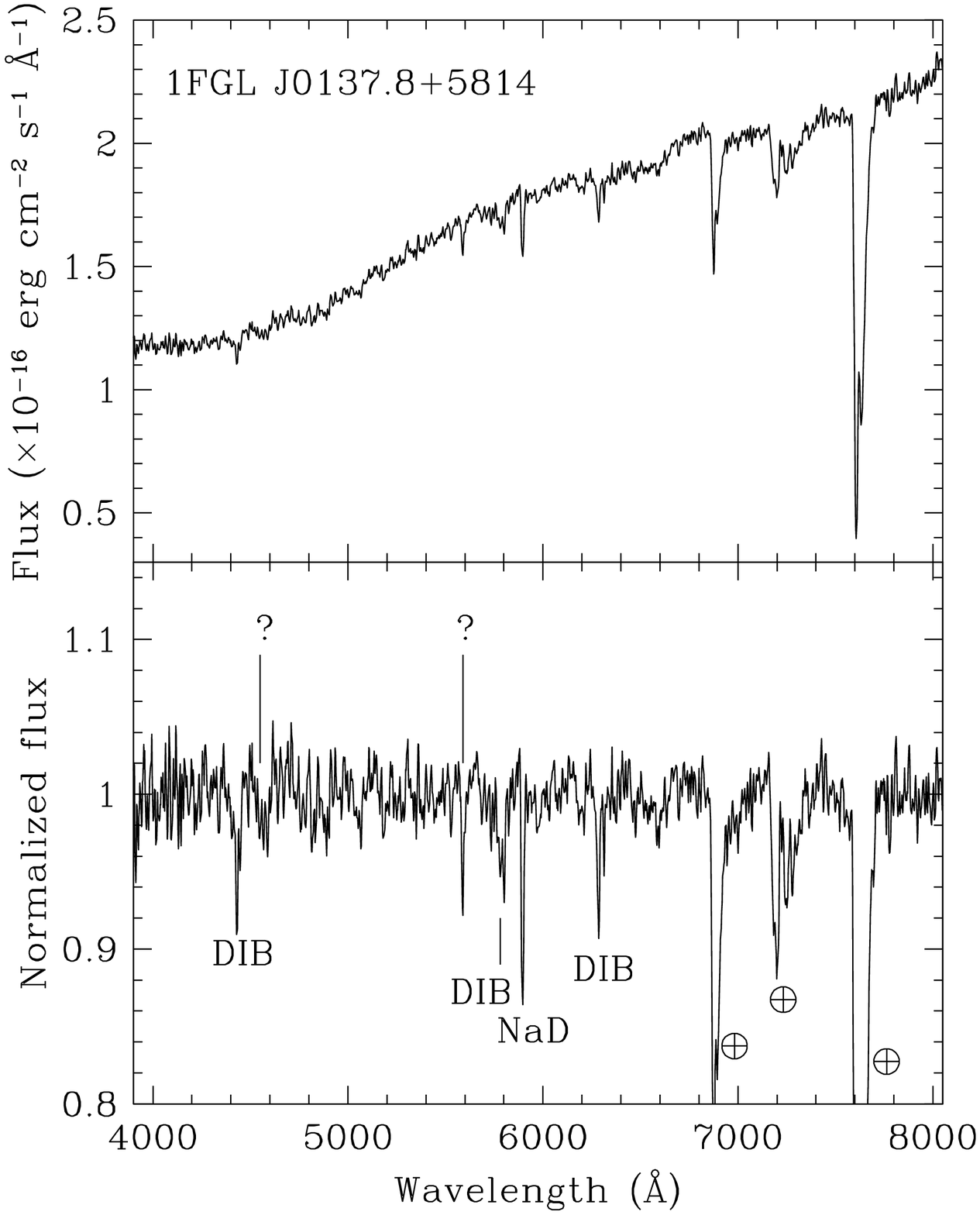,width=9cm,angle=0}}

\caption{Spectra of the optical counterparts of four BL Lacs in our 
sample. For each panel, the upper box reports the observed spectrum not 
corrected for the intervening Galactic absorption, while the lower one 
shows the same spectrum normalized to its continuum. The main spectral 
features are labeled in the lower panels, with the intervening Galactic 
features indicated in the lower part of the box and (if present) the 
features allowing the redshift measurement marked in the above part of 
it. The symbol $\oplus$ indicates atmospheric telluric absorption bands. 
The TNG spectra have been smoothed using a Gaussian filter with $\sigma$ = 
3 \AA.}
\end{figure*}

\begin{figure*}
\mbox{\psfig{file=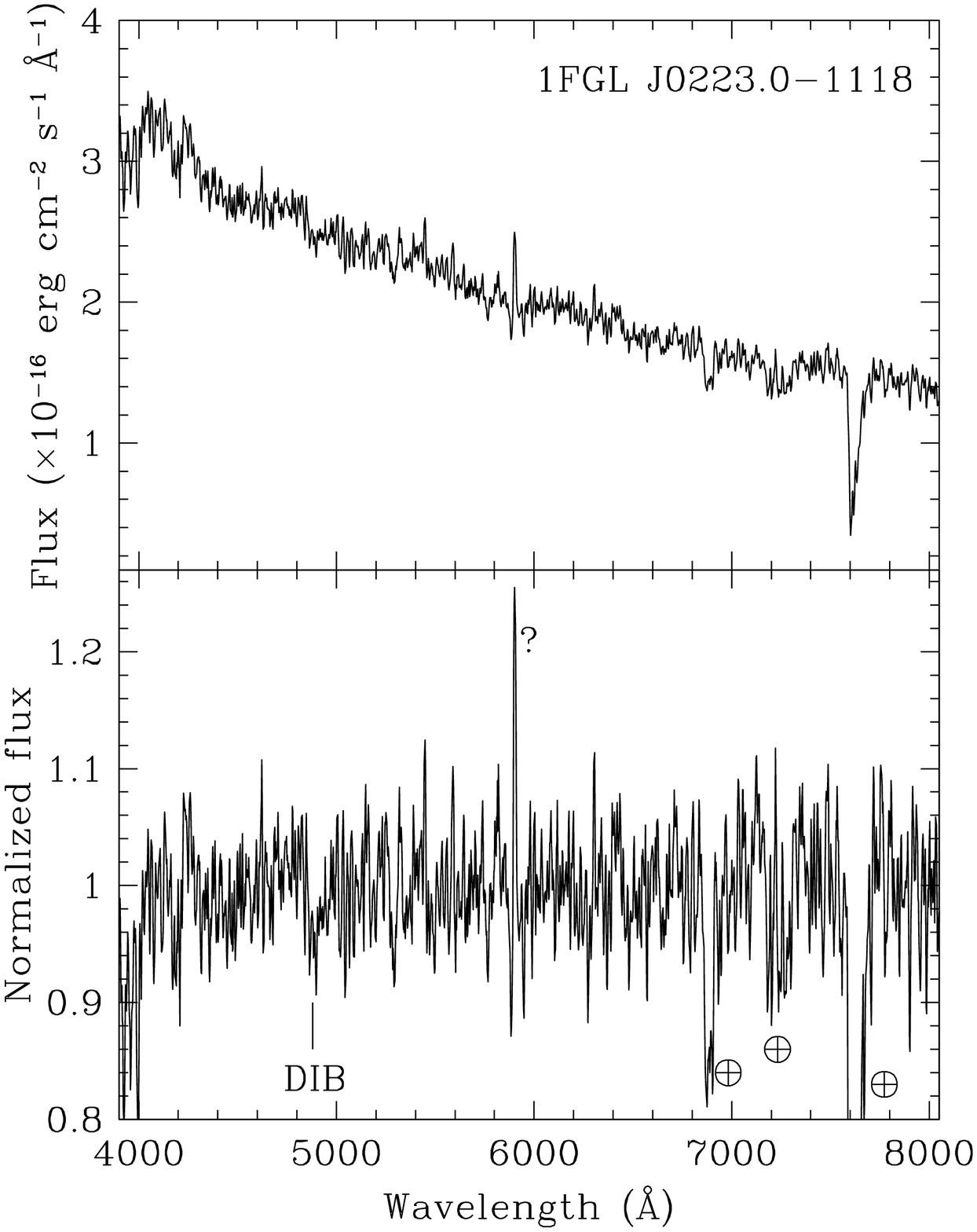,width=9cm,angle=0}}
\mbox{\psfig{file=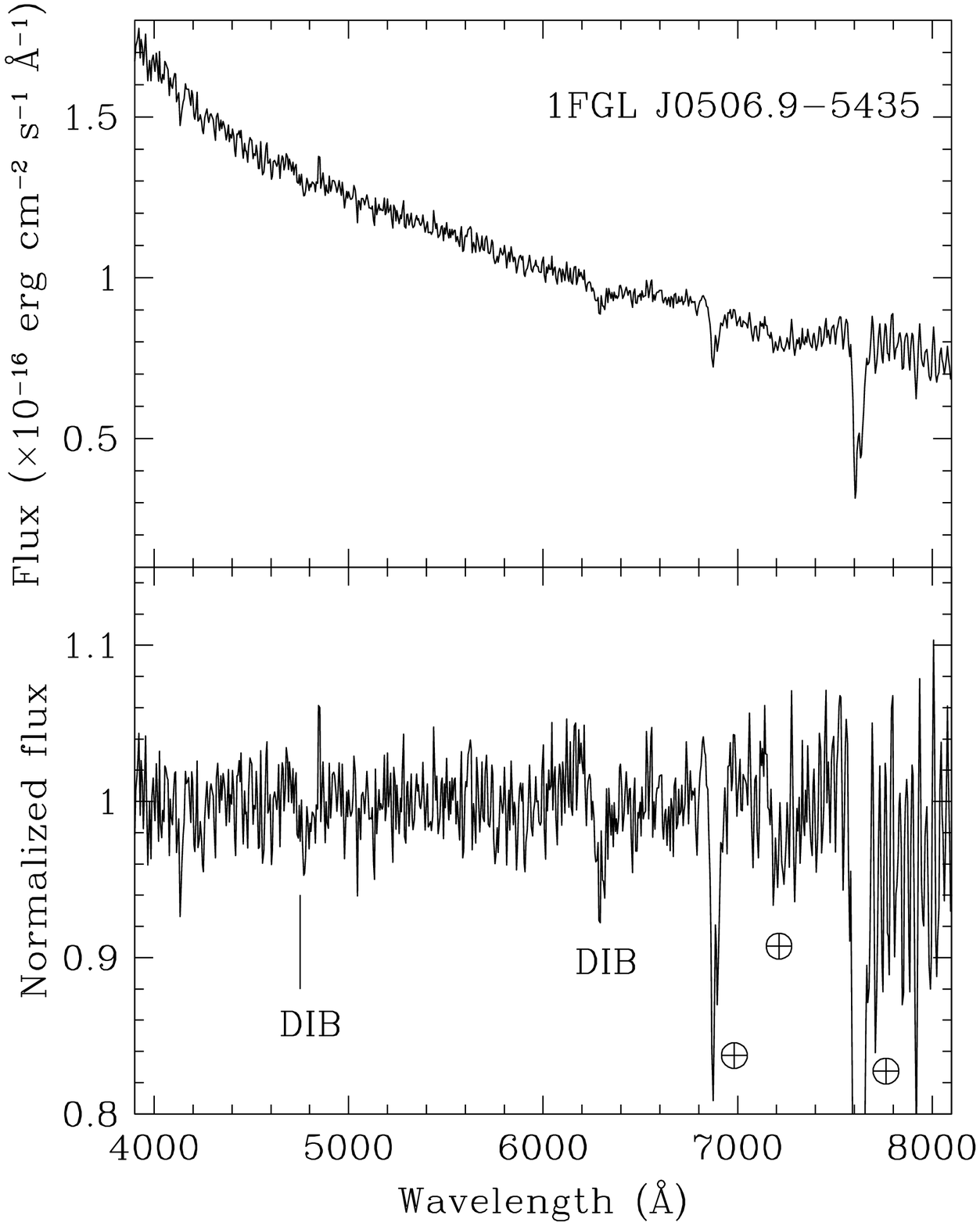,width=9cm,angle=0}}

\vspace{-.9cm}
\mbox{\psfig{file=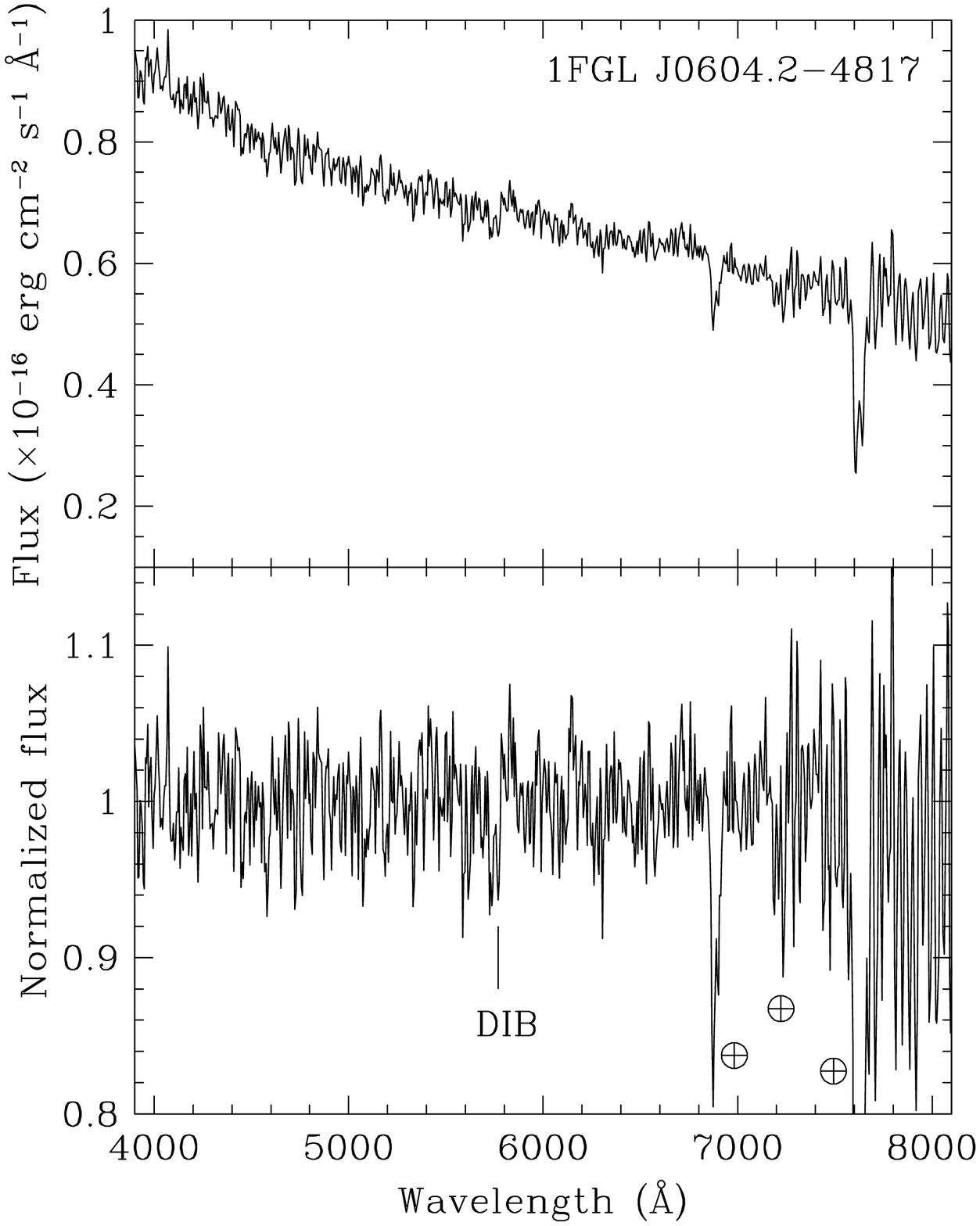,width=9cm,angle=0}}
\mbox{\psfig{file=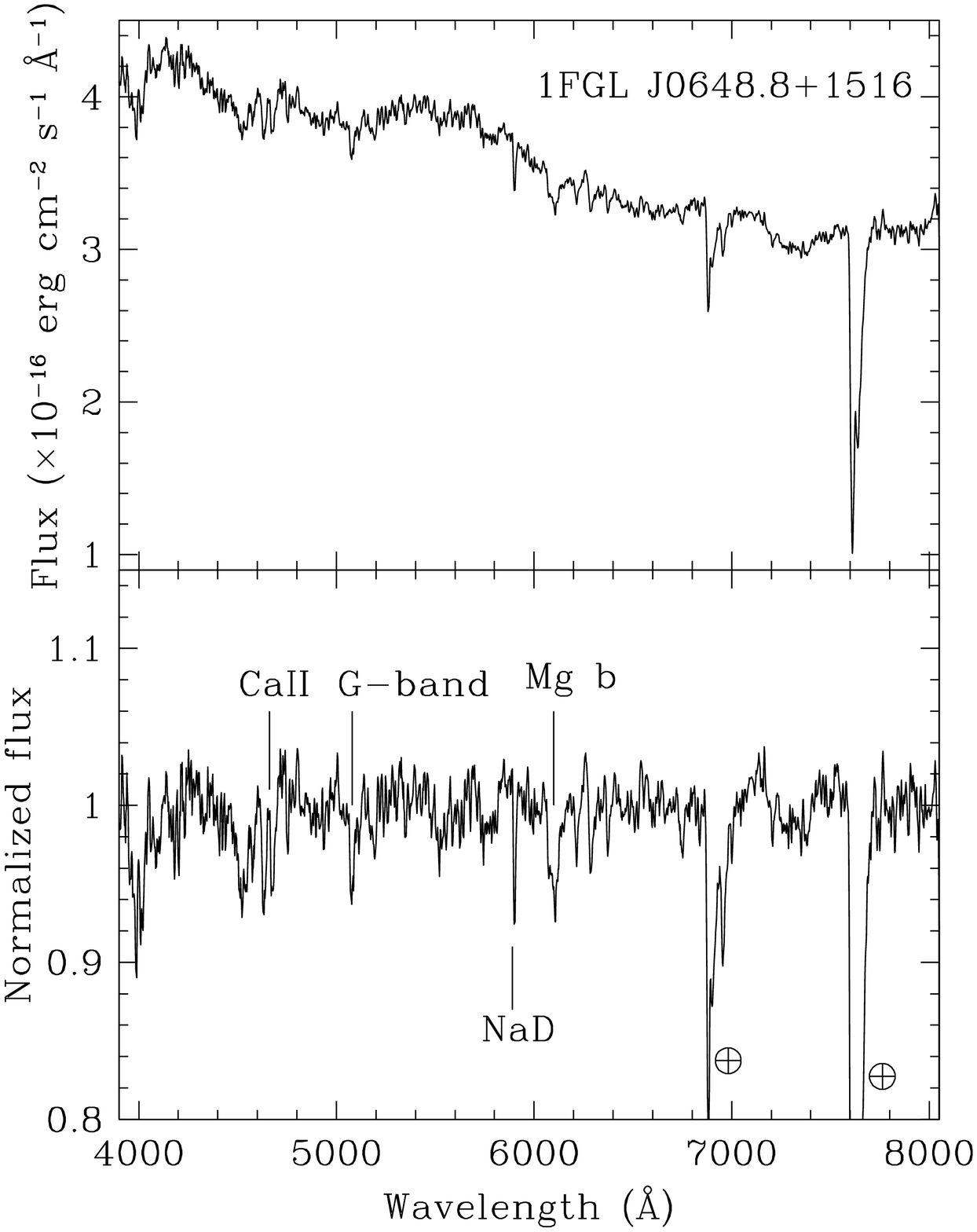,width=9cm,angle=0}}

\caption{The same as Fig. 1, but for four more spectra of BL Lacs in
our sample.}
\end{figure*}

\begin{figure*}
\mbox{\psfig{file=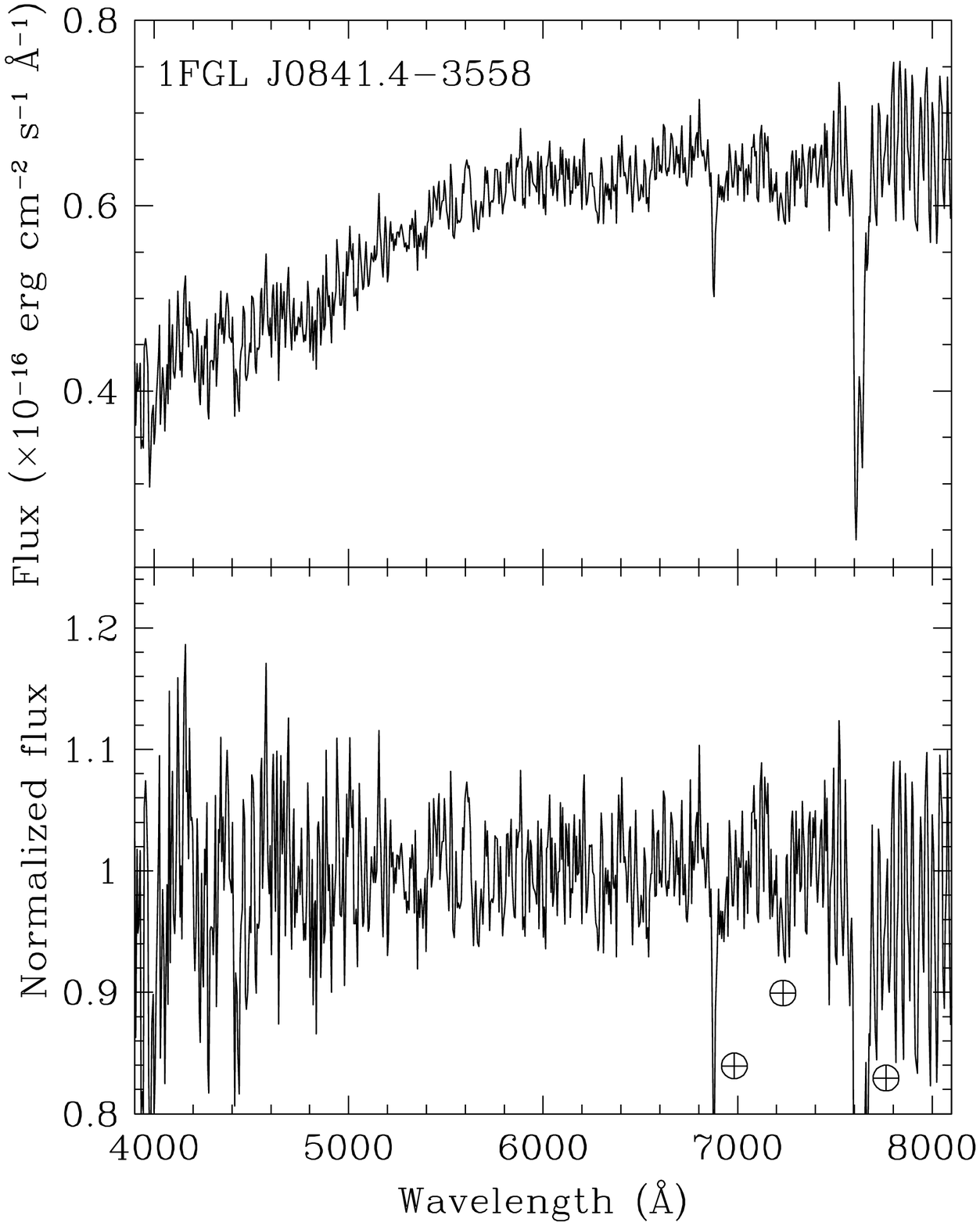,width=9cm,angle=0}}
\mbox{\psfig{file=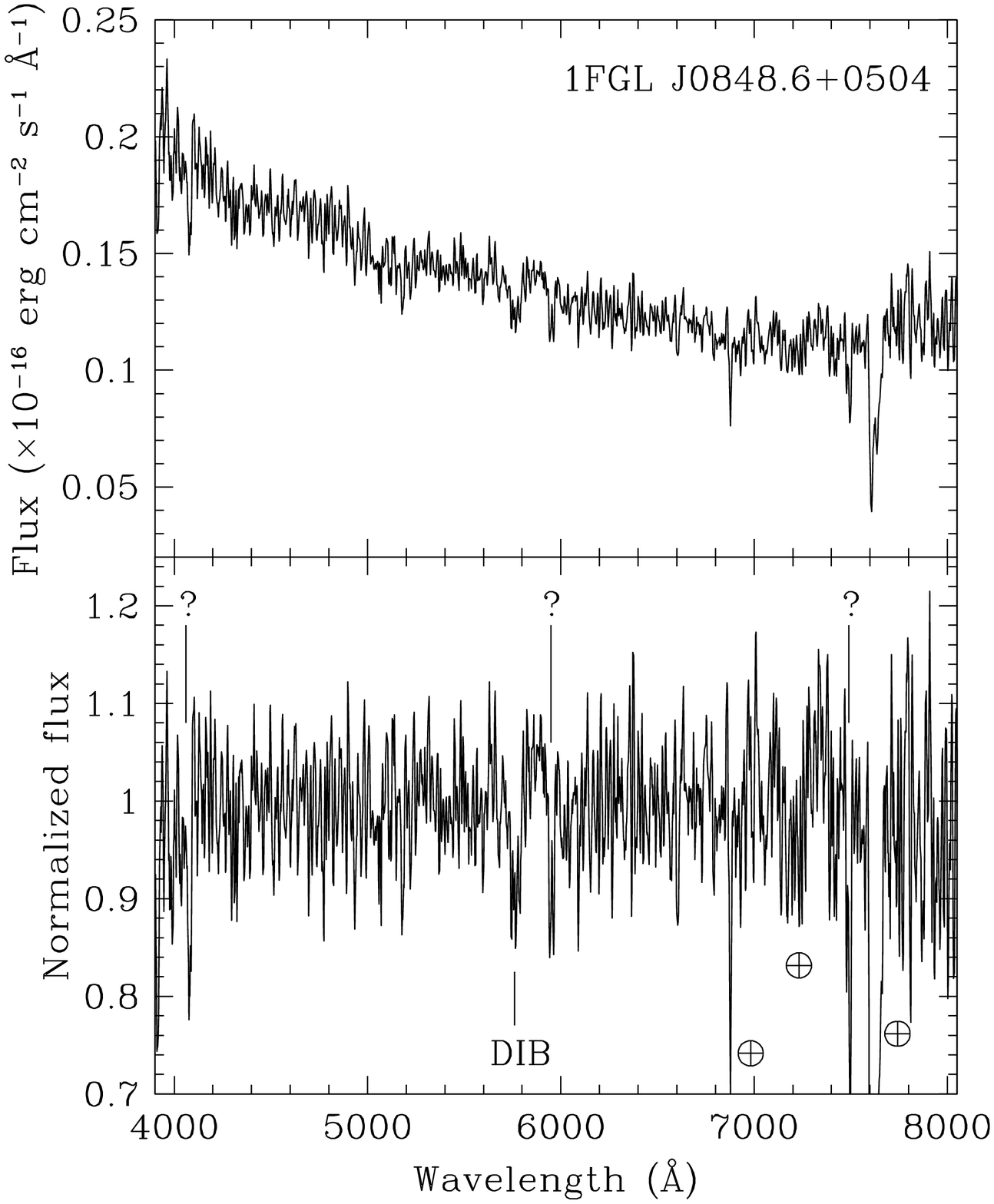,width=9cm,angle=0}}

\vspace{-.9cm}
\mbox{\psfig{file=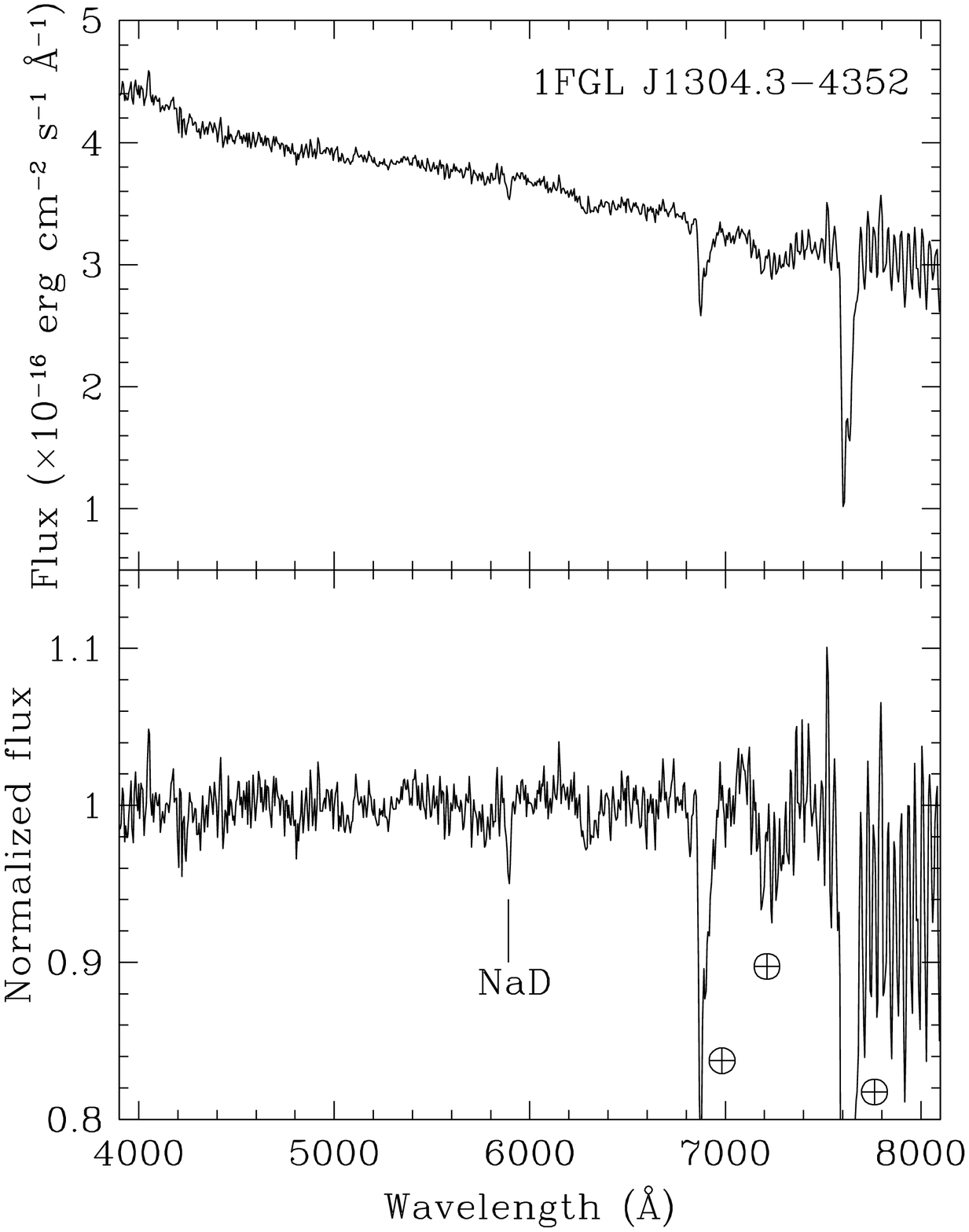,width=9cm,angle=0}}
\mbox{\psfig{file=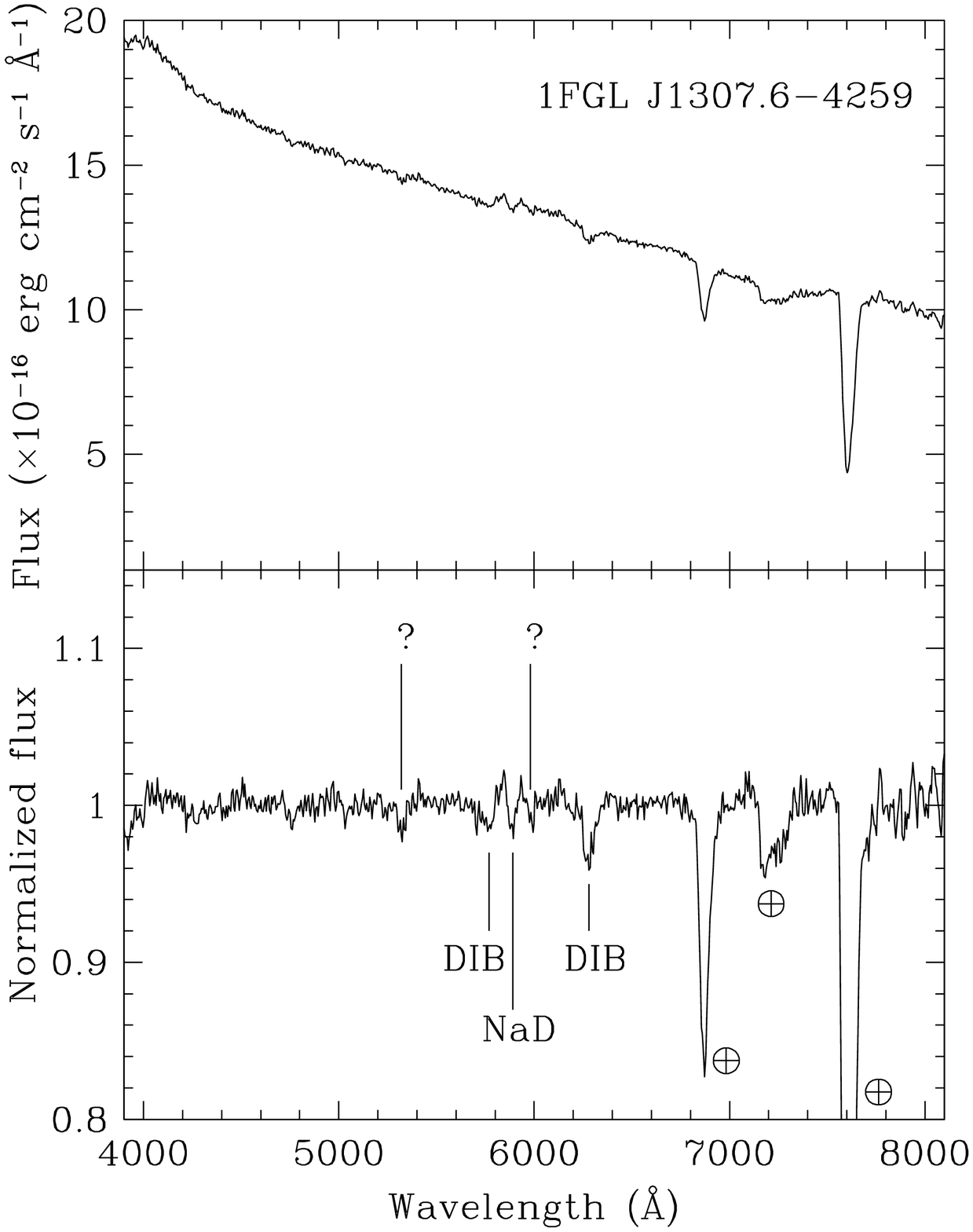,width=9cm,angle=0}}

\caption{The same as Fig. 1, but for four more spectra of BL Lacs in
our sample.}
\end{figure*}

\begin{figure*}
\mbox{\psfig{file=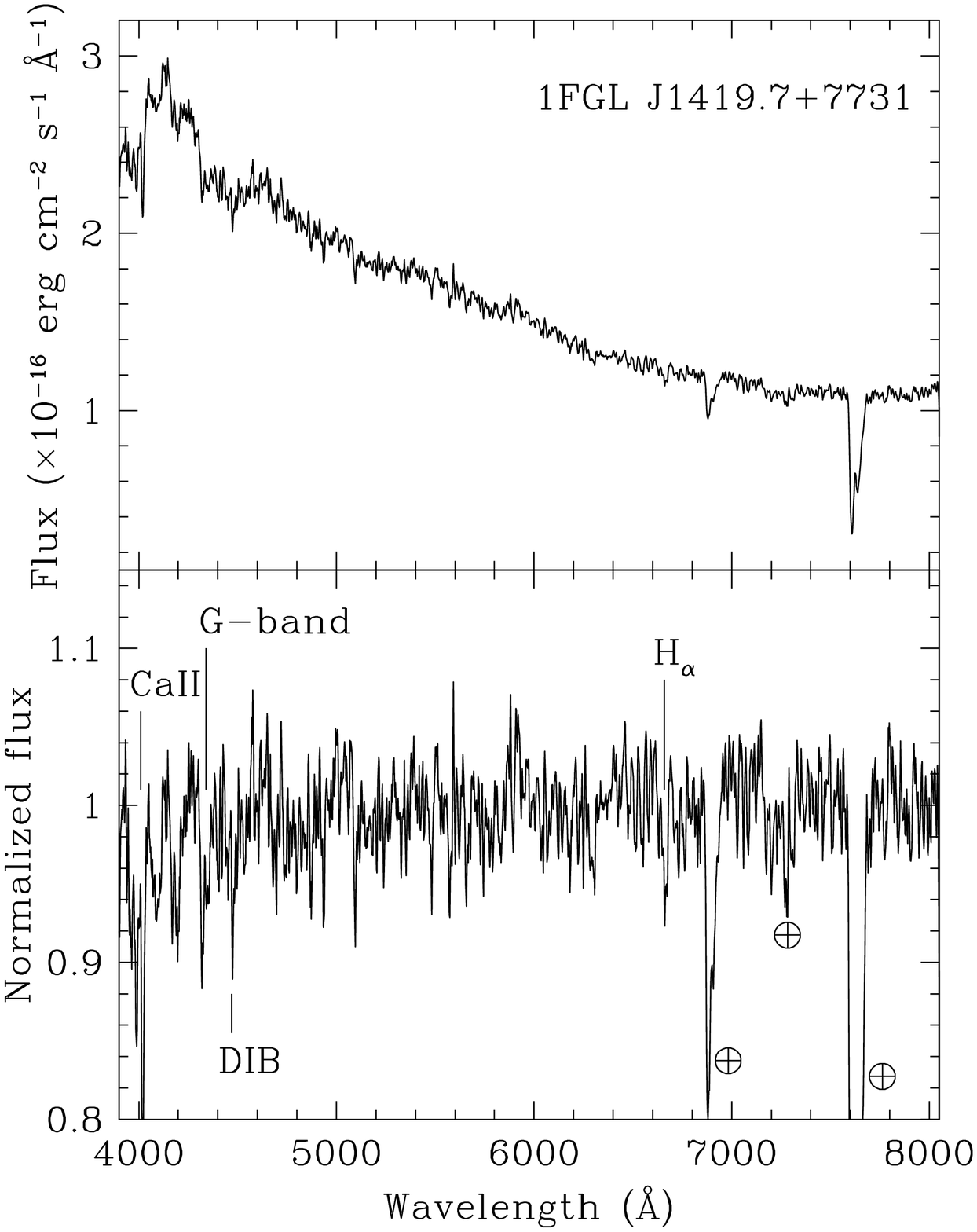,width=9cm,angle=0}}
\mbox{\psfig{file=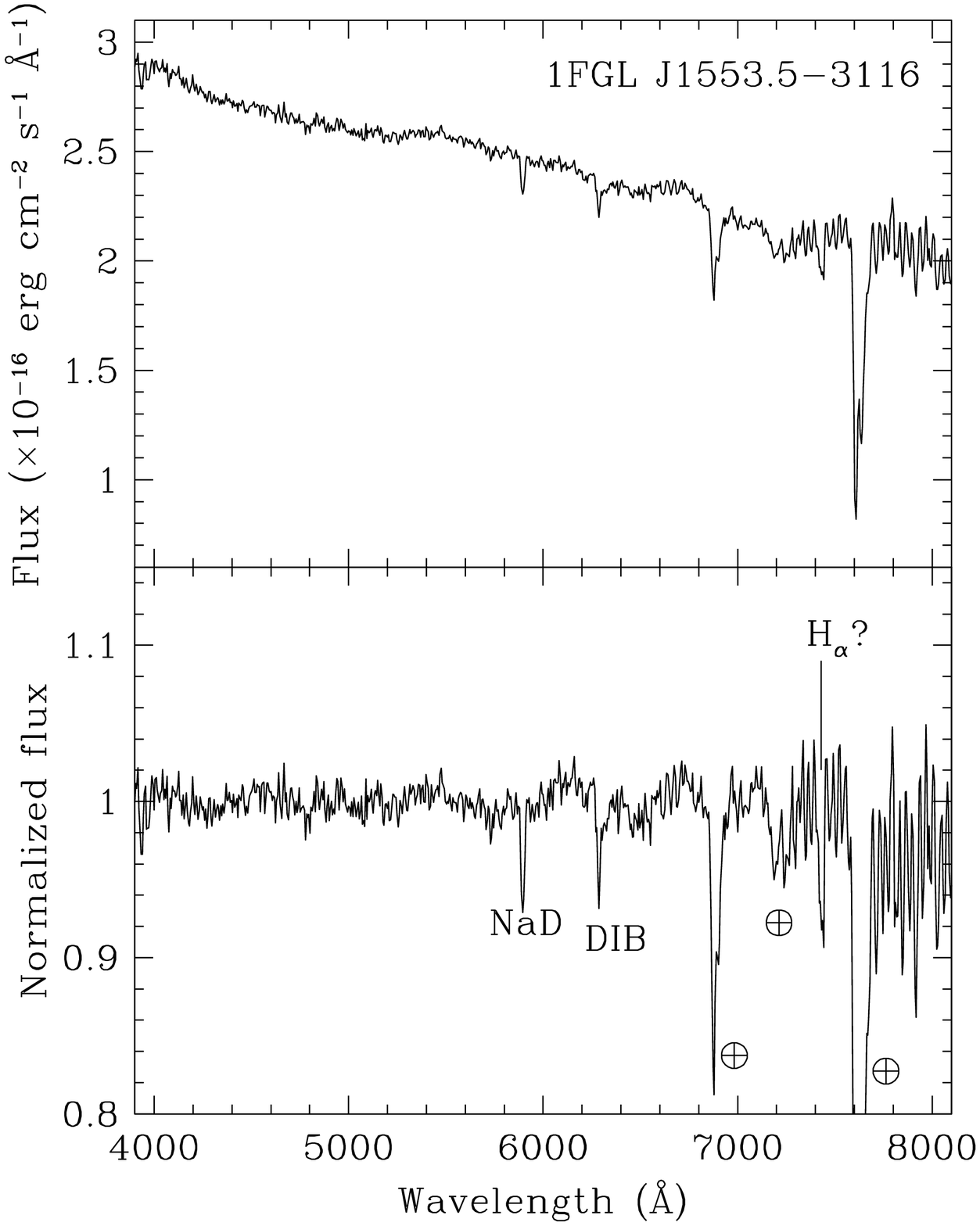,width=9cm,angle=0}}

\vspace{-.9cm}
\mbox{\psfig{file=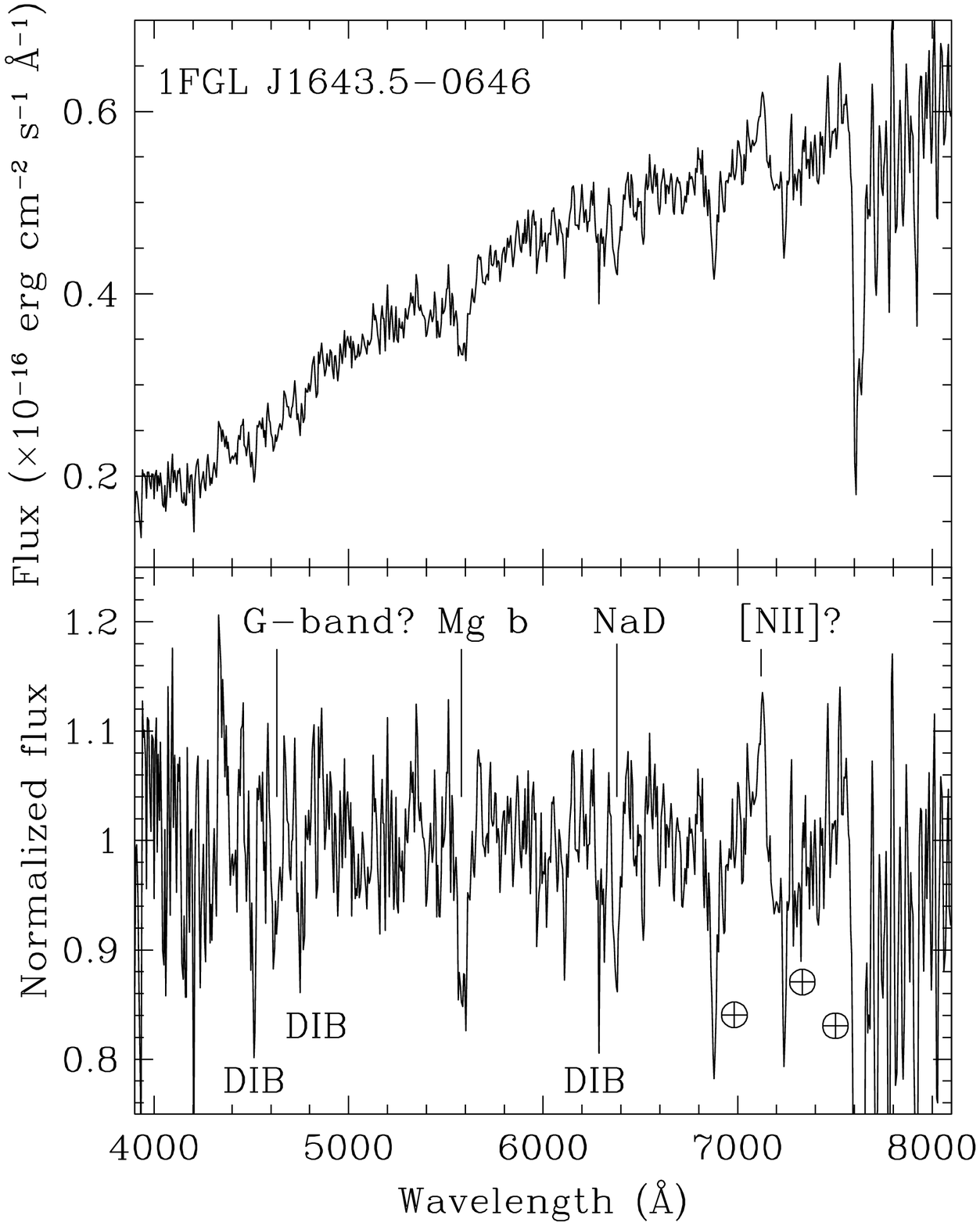,width=9cm,angle=0}}
\mbox{\psfig{file=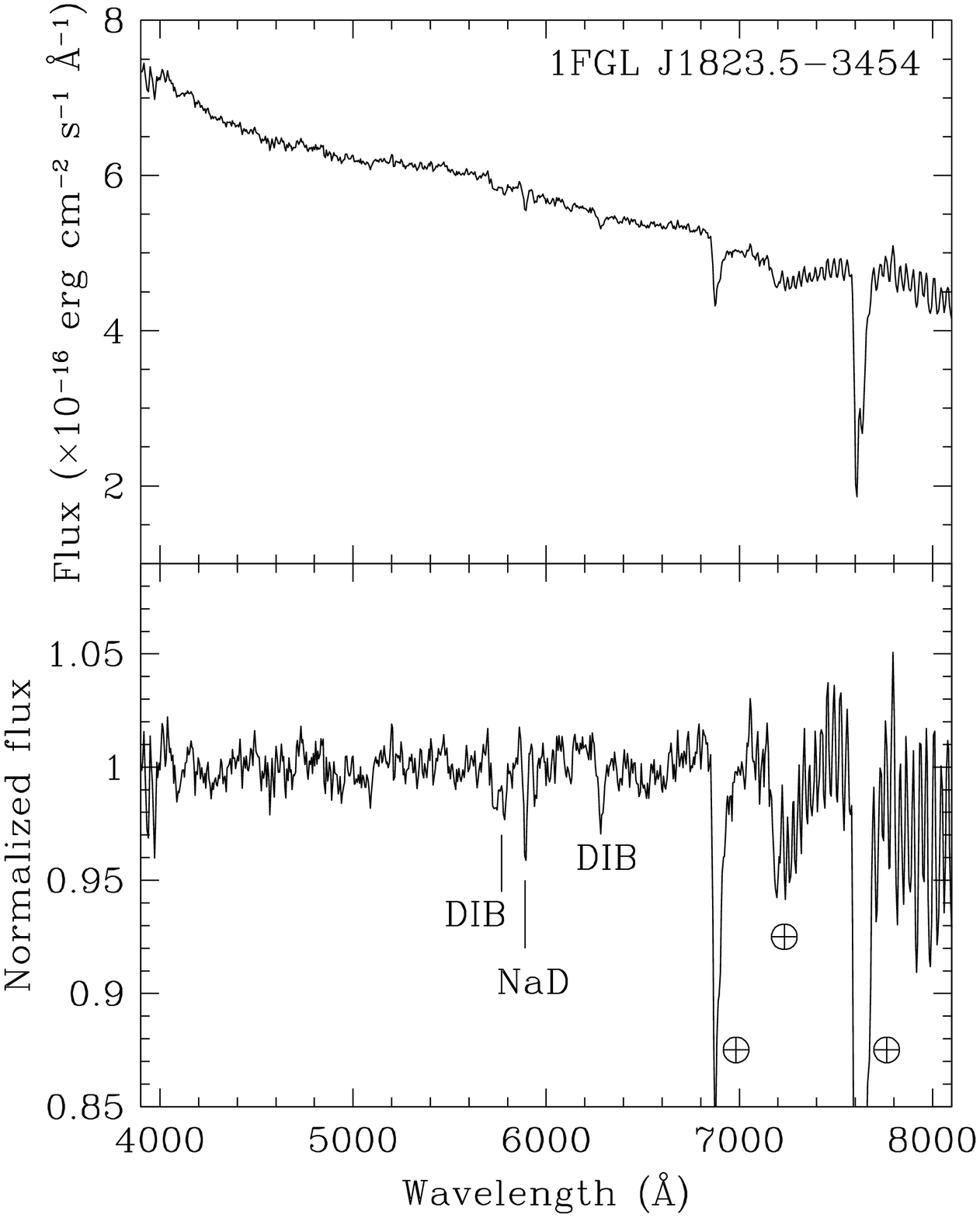,width=9cm,angle=0}}

\caption{The same as Fig. 1, but for four more spectra of BL Lacs in
our sample.}
\end{figure*}

\begin{figure*}
\mbox{\psfig{file=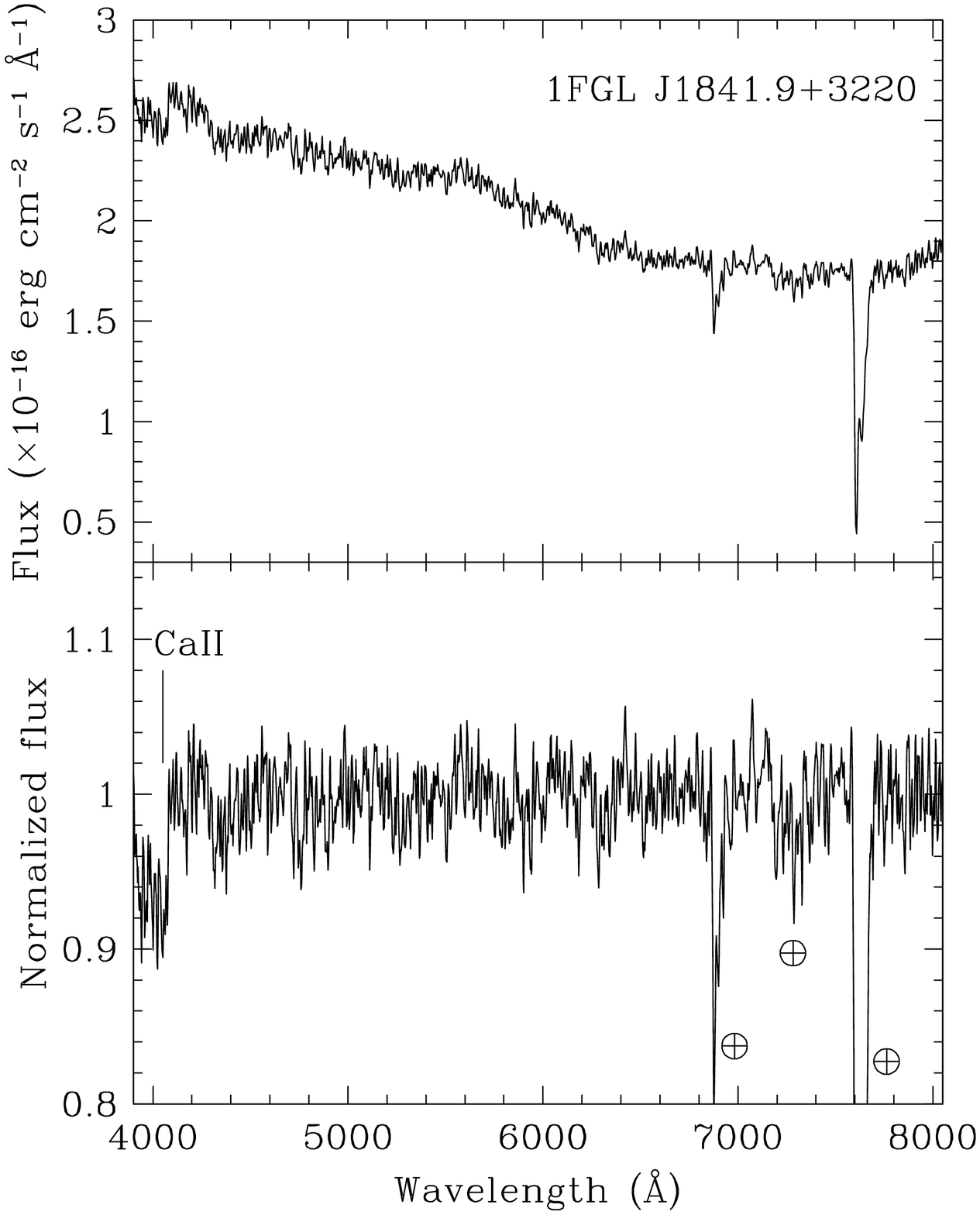,width=9cm,angle=0}}
\mbox{\psfig{file=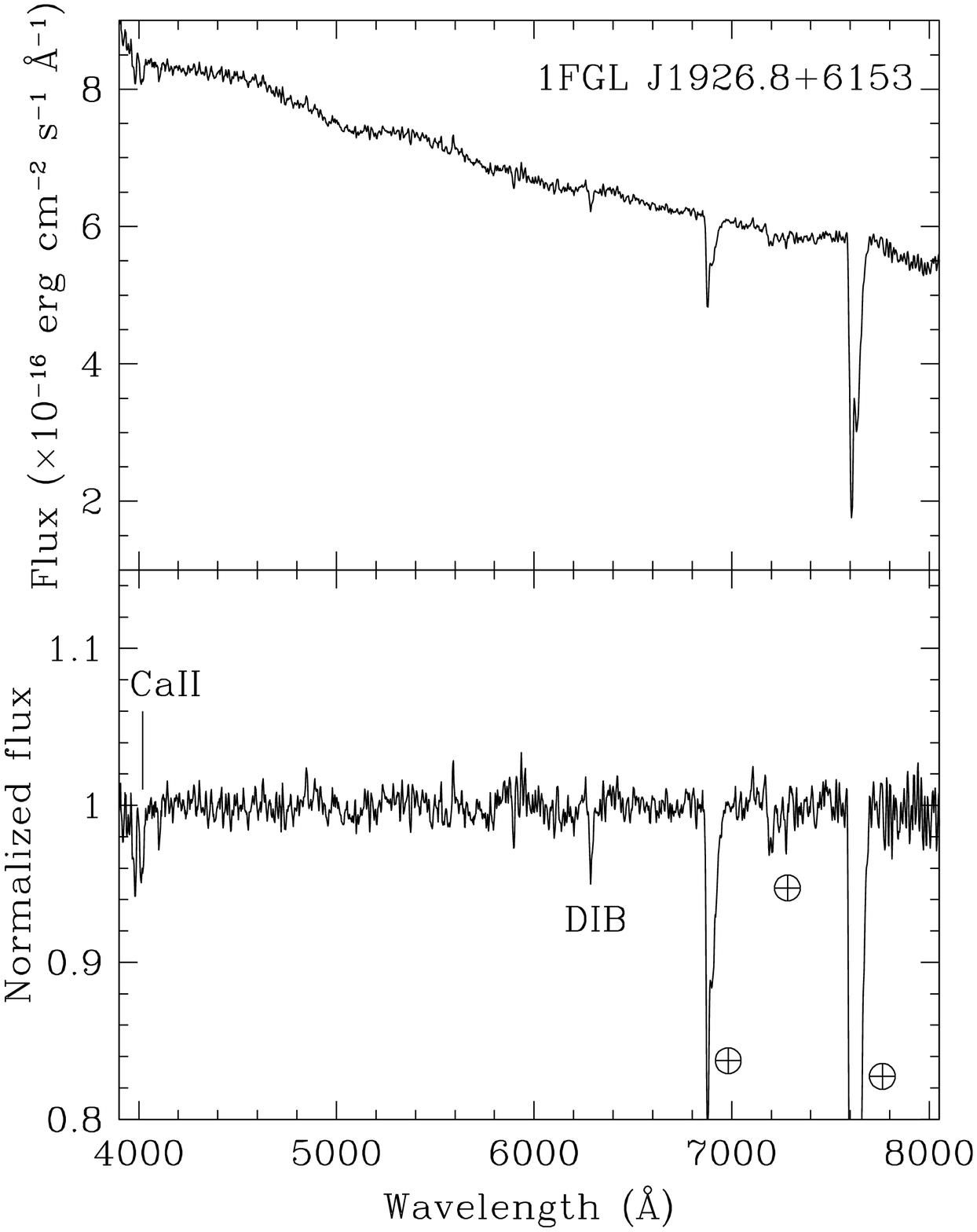,width=9cm,angle=0}}

\vspace{-.9cm}
\mbox{\psfig{file=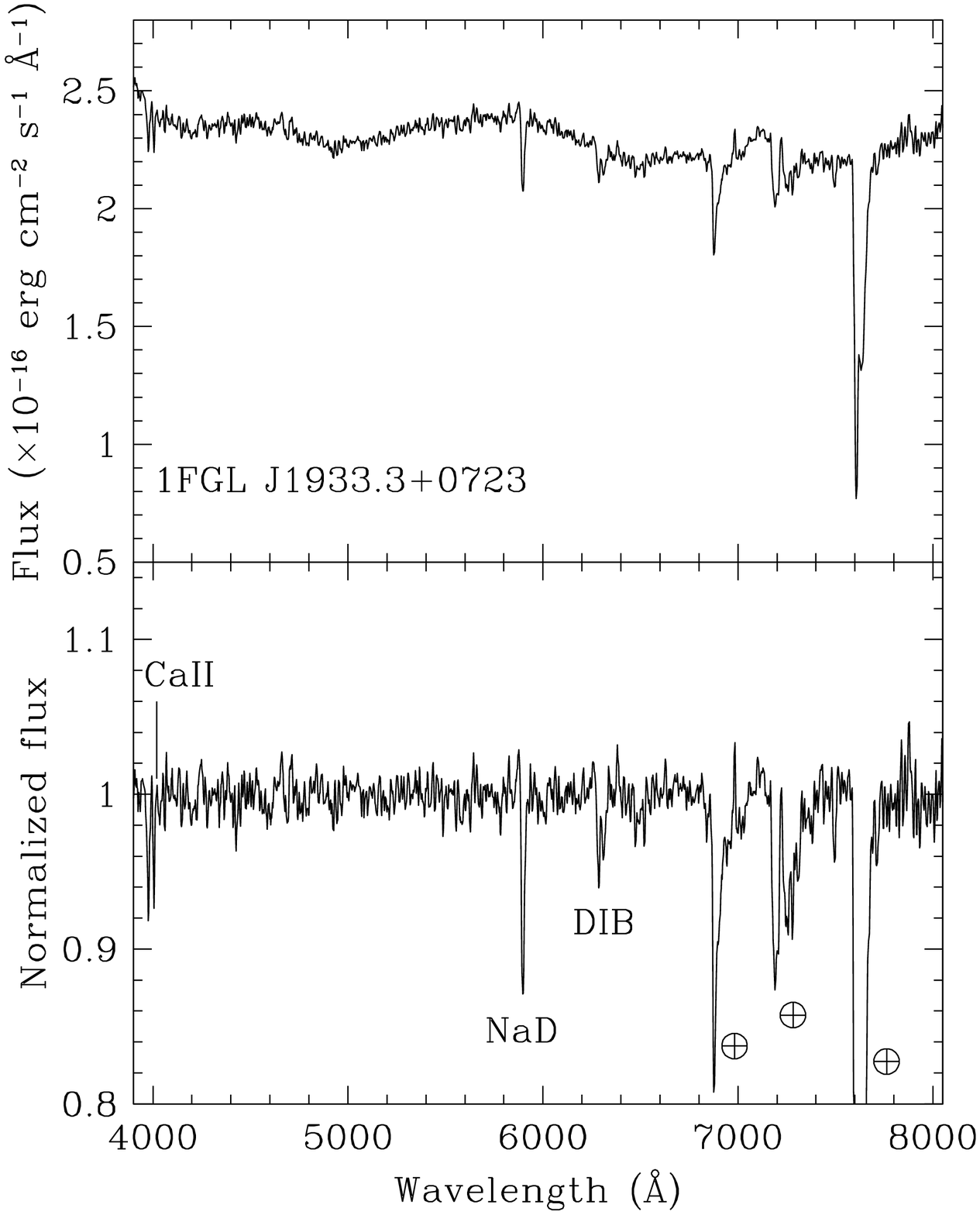,width=9cm,angle=0}}
\mbox{\psfig{file=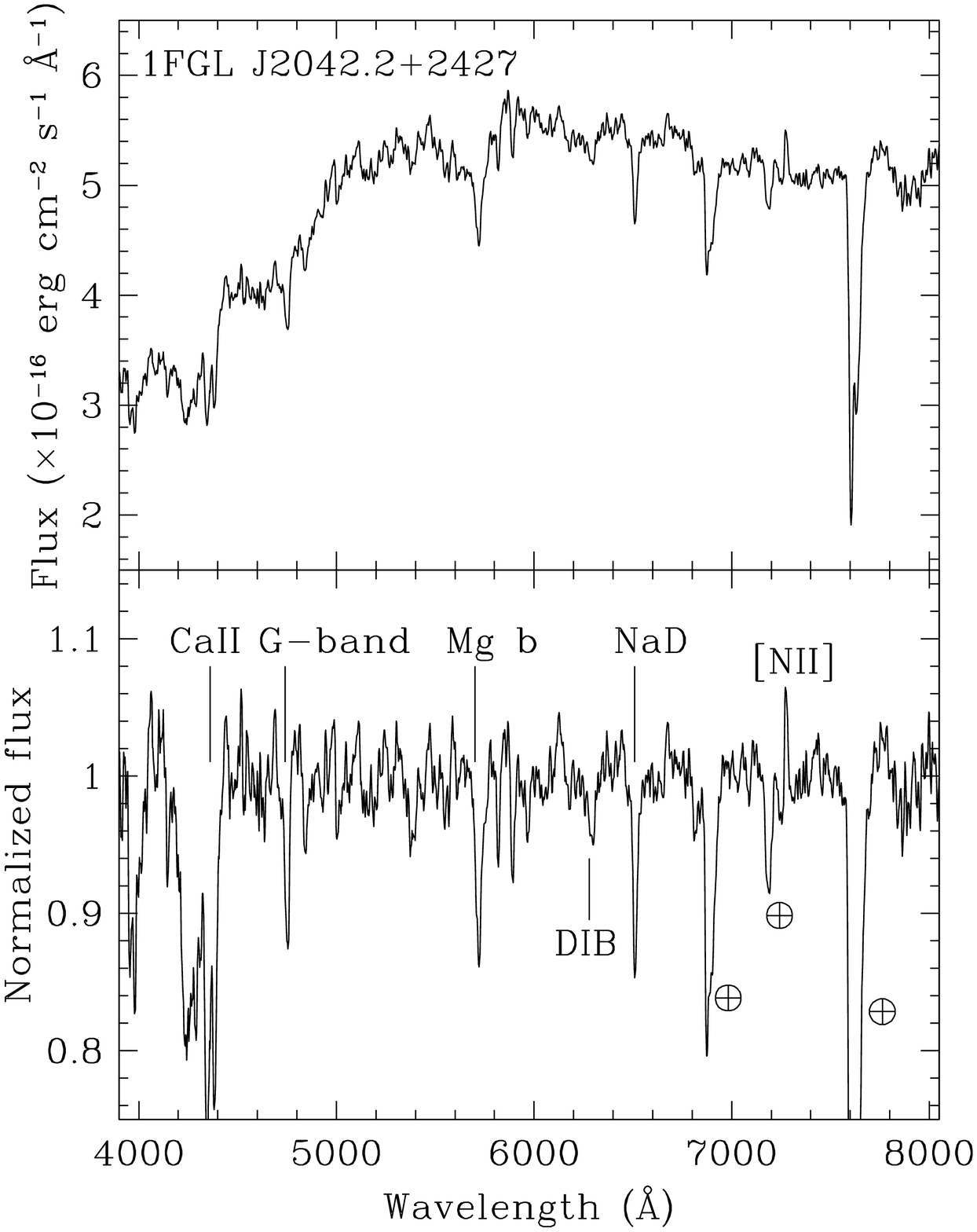,width=9cm,angle=0}}

\caption{The same as Fig. 1, but for four more spectra of BL Lacs in
our sample.}
\end{figure*}

\begin{figure*}[th!]
\mbox{\psfig{file=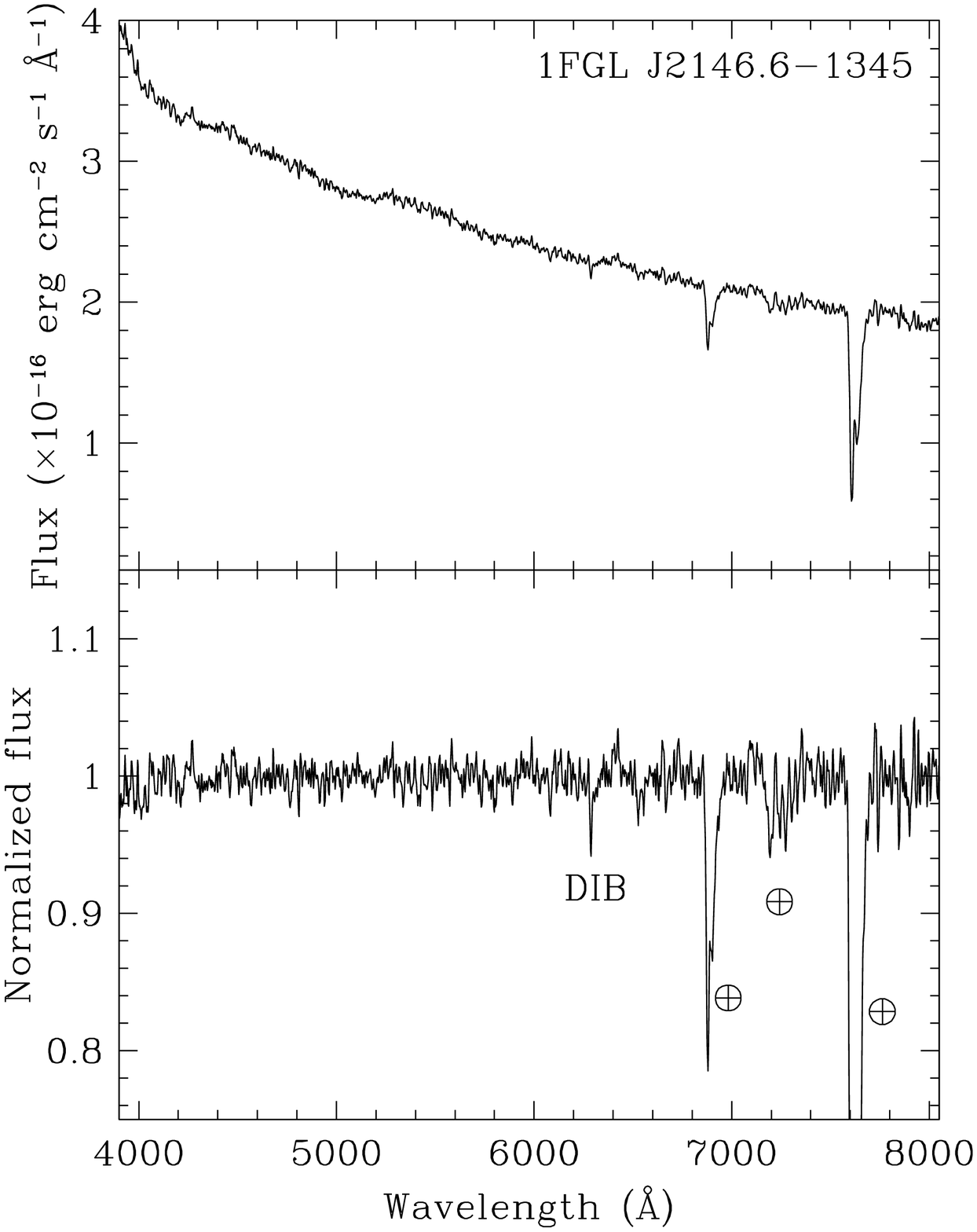,width=9cm,angle=0}}
\mbox{\psfig{file=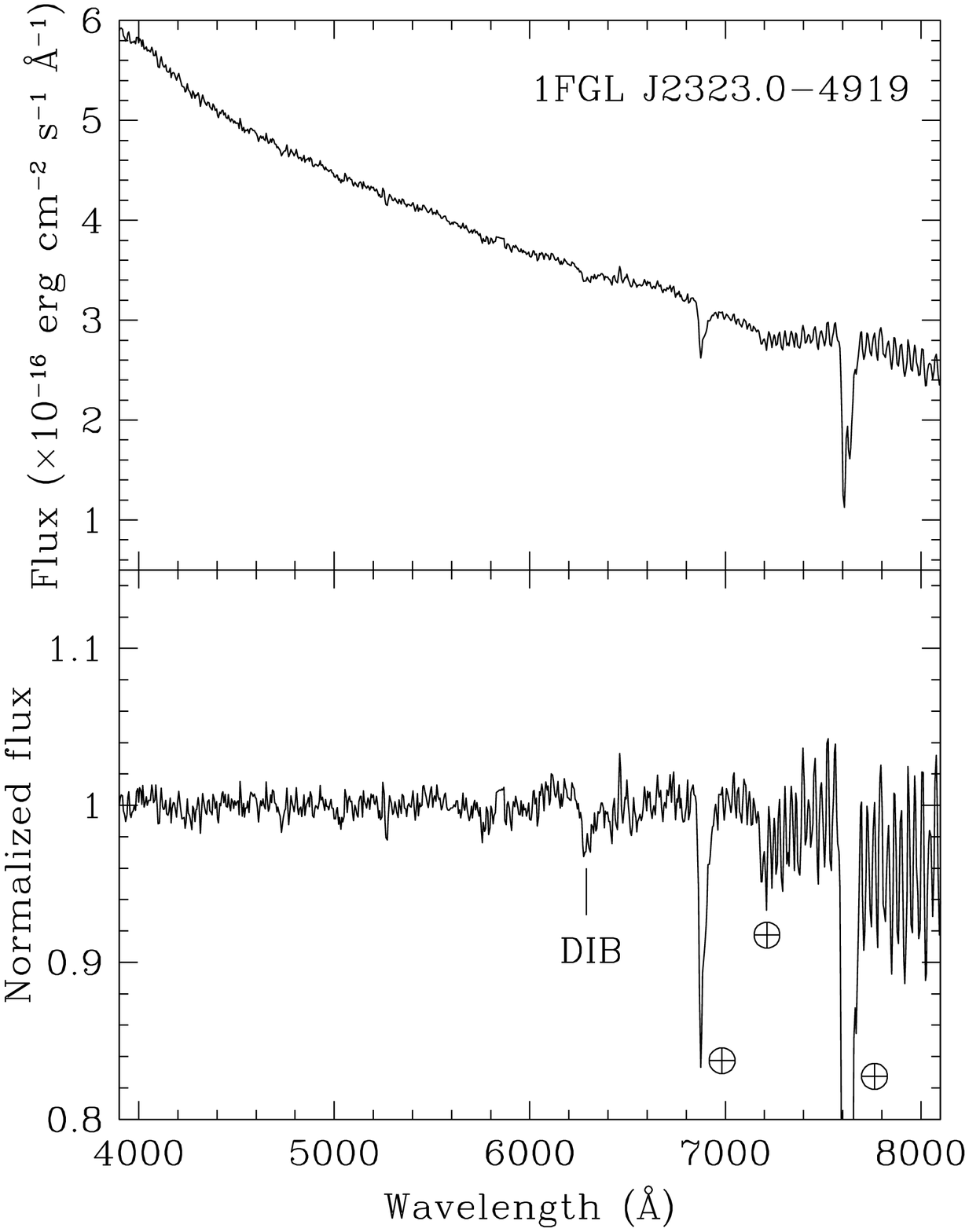,width=9cm,angle=0}}

\vspace{-.9cm}
\parbox{9.5cm}{
\psfig{file=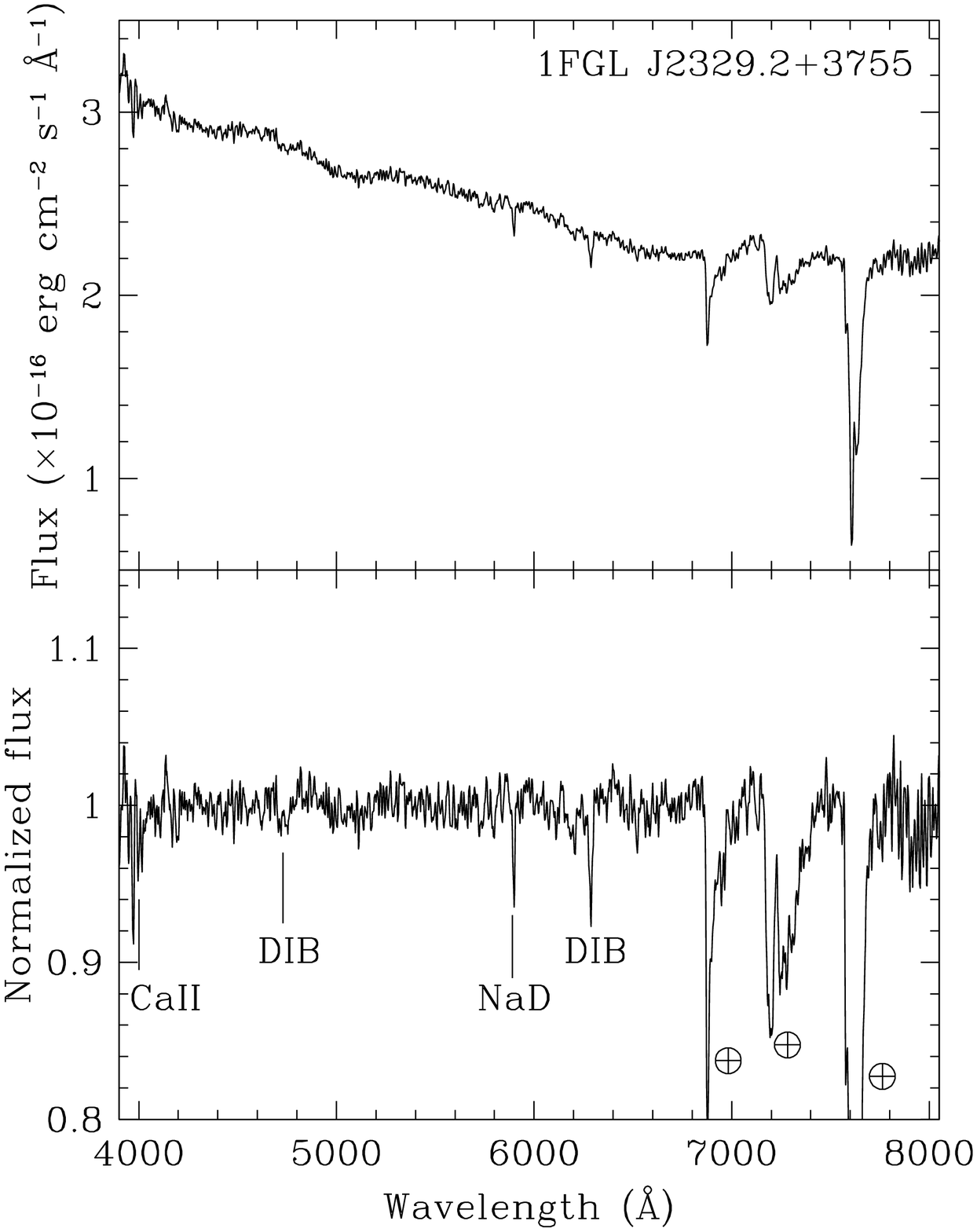,width=9cm,angle=0}
}
\hspace{0.8cm}
\parbox{8cm}{
\vspace{-.5cm}
\caption{The same as Fig. 1, but for three more spectra of BL Lacs in
our sample.}}
\end{figure*}

Our optical spectroscopy shows that almost all of the sources (23 cases; 
Figs. 1-6) in our sample are BL Lac-type active galactic nuclei (AGNs). 
Their spectra are mainly characterized by a nearly featureless and power 
law shaped continuum (see, e.g., Laurent-Muehleisen et al. 1998; see also 
Sect. 5), with the sole exception of 1FGL J2042.2+2427 (lower-right panel 
of Fig. 5). This makes the determination of absorption and/or emission 
lines difficult, as the contribution of the host galaxy to the total light 
is often overwhelmed by that of the AGN jet pointing towards the observer, 
as expected in BL Lacs.

Because of this, we also present in Figs. 1-6 the spectra normalized to 
the continuum; this is done to enhance the spectral contrast and to make 
the search for spectral features easier. We could determine the redshift 
for four objects through the identification of multiple features, whereas 
for four more BL Lacs we give a tentative redshift by means of the 
identification of a single absorption feature (either H$_\alpha$ or Ca{\sc 
ii}; see Table 3). In six out of these eight cases we cannot exclude that 
the observed absorption lines are produced by an intervening system 
because of the discrepancy with the lower limit of the object redshift in 
Table 2 (see below), as well as with those computed by Shaw et al. (2013). 
We note, however, that the measured redshifts of 1FGL J1643.5$-$0646 and 
1FGL J2042.2+2427 are actually smaller than the corresponding lower limits 
set by Shaw et al. (2013).

In two cases (1FGL J0838.6$-$2828 and 1FGL J1544.5$-$1127) we identified 
the {\it ROSAT} sources found by Stephen et al. (2010) within the {\it 
Fermi} error ellipse as Galactic cataclysmic variables (CVs). A thorough 
description of the results concerning these two sources is given in the 
Appendix.

We also report that the optical spectrum of the bright star HD 74208 
associated with the {\it ROSAT} source 1RXS J084121.4$-$355704, i.e., the 
softer of the two X--ray objects detected by {\it Swift}/XRT within the 
error box of 1FGL J0841.4$-$3558 (see Sect. 2), is indeed typical of an 
F-type star with no peculiarities. Therefore we will not discuss this 
object further and we will assume that the correct counterpart of 1FGL 
J0841.4$-$3558 is, also in this case, a BL Lac AGN (Fig. 3, upper-left 
panel).

It is worth noting that our results concerning 1FGL J0648.8+1516 (in 
Table 3) are consistent with those of Aliu et al. (2011); moreover, eight 
of the BL Lac objects in our sample (indicated with an asterisk in Table 
2) are also present in the large sample of BL Lac spectra published by 
Shaw et al. (2013).

Finally, for all BL Lac objects listed in Table 2 (including 1FGL 
J1353.6$-$6640 and 1FGL J1942.7+1033) we applied the method of Sbarufatti 
et al. (2006; see also Sbarufatti et al. 2009) to determine a lower limit 
for the source redshift and the optical continuum power law spectral 
slope $\alpha$; this is possible through the use of the value of the 
minimum equivalent width (EW) for an absorption line that can be detected 
in the available optical spectrum of the source itself. This approach can 
be performed through the equation

\begin{equation}
\mathrm{EW}_{\rm obs}=\frac{(1+z) \times \mathrm{EW}_0}{1+\rho/A(z)}~,
\end{equation}

\noindent
where EW$_{\rm obs}$ is the observed EW of an absorption line, EW$_0$ its 
intrinsic value, $z$ the source redshift, $\rho$ the nucleus-to-host flux 
ratio, and $A(z)$ a correction term that takes into account the loss of 
light inside the observed slit aperture. The results of this method are 
also listed in Table 2, together with the results of the photometry of the 
optical counterpart of 1FGL J1823.5$-$3454; the uncertainty on the latter 
measurement can be estimated as $\pm$0.3 mag, which is the systematic 
error on the USNO magnitude calibration (Monet et al. 2003), given that 
the statistical error of our photometry on this source is negligible 
compared to the systematic error of the USNO stars used as calibrators. 
Comparing our results of Table 2 with those of Shaw et al. (2013) for the 
objects belonging to both samples, we see that our lower limit for the 
redshift of a given object is in general lower than that of those authors.

\begin{table*}[th!]
\caption[]{List of parameters corresponding to the BL Lac objects in our 
sample obtained with the approach of Sbarufatti et al. (2006, 2009; see 
text).}
\begin{center}
\begin{tabular}{llccccc}
\noalign{\smallskip}
\hline
\hline
\noalign{\smallskip}
\multicolumn{1}{c}{Object} & \multicolumn{1}{c}{$R$} & $E(B-V)_{\rm Gal}$ & $\alpha$ & EW$_{\rm min}$ & Redshift & Log($\nu_{\rm peak}$) \\
 & \multicolumn{1}{c}{mag.} & (mag.) & & (\AA) & & (Hz) \\

\noalign{\smallskip}
\hline
\noalign{\smallskip}

1FGL J0051.4$-$6242     & 17.3           & 0.016 & $-$1.0$\pm$0.1 & 0.3 & $>$0.30 & 15.06 \\ 
1FGL J0054.9$-$2455$^*$ & 17.1           & 0.016 & $-$1.3$\pm$0.3 & 1.2 & $>$0.12 & 15.10 \\ 
1FGL J0131.2+6121       & 18.2           & 0.922 & $-$1.5$\pm$0.4 & 1.3 & $>$0.08 & 15.85 \\ 
1FGL J0137.8+5814       & 17.5           & 0.528 & $-$0.7$\pm$0.2 & 0.7 & $>$0.12 & 14.98 \\ 
1FGL J0223.0$-$1118     & 17.4           & 0.021 & $-$1.3$\pm$0.4 & 1.6 & $>$0.20 & 14.84 \\ 
1FGL J0506.9$-$5435     & 16.1           & 0.030 & $-$1.2$\pm$0.2 & 0.7 & $>$0.40 & 15.61 \\ 
1FGL J0604.2$-$4817     & 17.4           & 0.050 & $-$0.9$\pm$0.2 & 1.0 & $>$0.37 & 15.20 \\ 
1FGL J0648.8+1516       & 16.2           & 0.144 & $-$1.0$\pm$0.2 & 0.9 & $>$0.13 & 15.52 \\ 
1FGL J0841.4$-$3558     & 17.2           & 0.495 & $-$0.9$\pm$0.3 & 1.8 & $>$0.15 & 15.01 \\ 
1FGL J0848.6+0504       & 18.3           & 0.067 & $-$1.2$\pm$0.4 & 1.9 & $>$0.57 & 16.14 \\ 
1FGL J1304.3$-$4352$^*$ & 15.4           & 0.133 & $-$1.0$\pm$0.2 & 0.5 & $>$0.19 & 15.02 \\ 
1FGL J1307.6$-$4259     & 15.1           & 0.106 & $-$1.3$\pm$0.2 & 0.4 & $>$0.10 & 15.10 \\ 
1FGL J1353.6$-$6640     & 17.1           & 0.553 & $-$0.9$\pm$0.3 & 0.3 & $>$0.15 & 15.11 \\ 
1FGL J1419.7+7731       & 17.6           & 0.035 & $-$1.6$\pm$0.3 & 1.1 & $>$0.27 & 15.39 \\ 
1FGL J1553.5$-$3116$^*$ & 15.5           & 0.174 & $-$1.1$\pm$0.2 & 0.3 & $>$0.28 & 14.42 \\ 
1FGL J1643.5$-$0646$^*$ & 16.7           & 0.491 & +0.1$\pm$0.4   & 2.6 & $>$0.12 & 14.88 \\ 
1FGL J1823.5$-$3454     & 15.2$^{\rm a}$ & 0.129 & $-$1.1$\pm$0.2 & 0.4 & $>$0.17 & 15.50 \\ 
1FGL J1841.9+3220       & 17.0           & 0.088 & $-$1.0$\pm$0.2 & 0.8 & $>$0.20 & 15.38 \\ 
1FGL J1926.8+6153$^*$   & 16.6           & 0.058 & $-$0.9$\pm$0.2 & 0.4 & $>$0.16 & 15.29 \\ 
1FGL J1933.3+0723       & 15.3           & 0.295 & $-$1.2$\pm$0.2 & 0.4 & $>$0.20 & 14.64 \\ 
1FGL J1942.7+1033       & 17.0           & 0.390 & $-$0.1$\pm$0.2 & 0.2 & $>$0.20 & 14.75 \\ 
1FGL J2042.2+2427$^*$   & 14.0           & 0.140 & $^{**}$        & 1.9 & $^{**}$ & 15.32 \\ 
1FGL J2146.6$-$1345$^*$ & 17.5           & 0.042 & $-$1.1$\pm$0.2 & 0.5 & $>$0.28 & 14.92 \\ 
1FGL J2323.0$-$4919     & 17.1           & 0.012 & $-$1.1$\pm$0.2 & 0.4 & $>$0.27 & 15.04 \\ 
1FGL J2329.2+3755$^*$   & 17.0           & 0.182 & $-$1.3$\pm$0.2 & 0.6 & $>$0.20 & 15.15 \\ 

\noalign{\smallskip} 
\hline
\noalign{\smallskip} 

\multicolumn{7}{l}{Note: $R$-band magnitudes were obtained from the USNO-A2.0 catalog,} \\
\multicolumn{7}{l}{while the values of $E(B-V)_{\rm Gal}$ are from Schlegel et al. (1998).} \\
\multicolumn{7}{l}{$^{\rm a}$: obtained from the photometry on the acquisition image} \\
\multicolumn{7}{l}{of the spectrum presented in this work (see text).} \\
\multicolumn{7}{l}{$^*$: present in the BL Lac spectroscopic sample of Shaw et al. (2013).} \\
\multicolumn{7}{l}{$^{**}$: the optical emission of the source is dominated by the host galaxy;} \\
\multicolumn{7}{l}{thus, in this case it was not possible to fit the spectrum with a power law} \\
\multicolumn{7}{l}{according to the method of Sbarufatti et al. (2006).} \\

\noalign{\smallskip}
\hline
\hline
\noalign{\smallskip}
\end{tabular}
\end{center}
\end{table*}

\begin{table*}
\caption[]{Main optical spectroscopic information concerning BL Lac 
objects in our sample with detected redshifted spectral lines.}
\begin{center}
\begin{tabular}{lclccccc}
\noalign{\smallskip}
\hline
\hline
\noalign{\smallskip}
\multicolumn{1}{c}{Object} & Average & \multicolumn{1}{c}{Line} & $\lambda_{\rm obs}$ & Redshift & Type$^{\rm *}$ & EW & Observed \\
 & redshift & & (\AA) & & & (\AA) & flux \\

\noalign{\smallskip}
\hline
\noalign{\smallskip}

1FGL J0648.8+1516   & 0.179 &               &      &       &      &              &                \\
                    &       & Ca{\sc ii} H  & 4634 & 0.178 & $g$  &  1.8$\pm$0.4 & $-$7.0$\pm$1.4 \\
                    &       & Ca{\sc ii} K  & 4678 & 0.179 & $g$  &  1.8$\pm$0.4 & $-$7.4$\pm$1.5 \\
                    &       & G-band        & 5080 & 0.180 & $g$  &  1.9$\pm$0.4 & $-$7.1$\pm$1.4 \\
                    &       & Mg b          & 6104 & 0.180 & $g$  &  3.7$\pm$0.6 & $-$13$\pm$2     \\

 & & & & & & & \\

1FGL J1419.7+7731   & 0.014 &               &      &       &      &              &                \\
                    &       & Ca{\sc ii} H  & 3988 & 0.014 & $a$? &  1.2$\pm$0.4 & $-$2.4$\pm$0.8 \\
                    &       & Ca{\sc ii} K  & 4020 & 0.013 & $a$? &  2.7$\pm$0.6 & $-$7.0$\pm$1.5 \\
                    &       & G-band        & 4364 & 0.014 & $a$? &  3.8$\pm$0.8 & $-$10$\pm$2    \\
                    &       & H$_\alpha$    & 6662 & 0.015 & $a$? &  1.4$\pm$0.4 & $-$1.8$\pm$0.5 \\

 & & & & & & & \\

1FGL J1553.5$-$3116 & 0.132 &               &      &       &      &              &                \\
                    &       & H$_\alpha$?   & 7431 & 0.132 & $a$? &  3.8$\pm$0.8 & $-$8.2$\pm$1.7 \\

 & & & & & & & \\

1FGL J1643.5$-$0646 & 0.081 &               &      &       &      &              &                \\
                    &       & G-band        & 4645 & 0.079 & $g$  &  7.7$\pm$1.9 & $-$2.1$\pm$0.4 \\
                    &       & Mg b          & 5586 & 0.079 & $g$  & 12.9$\pm$1.3 & $-$5.3$\pm$0.5 \\
                    &       & Na D          & 6377 & 0.082 & $g$  &  8.5$\pm$1.3 & $-$4.4$\pm$0.7 \\
                    &       & [N {\sc ii}]? & 7124 & 0.082 & $e$  & $-$10$\pm$2  & 5.3$\pm$1.1    \\

 & & & & & & & \\

1FGL J1841.9+3220   & 0.028 &               &      &       &      &             &                \\
                    &       & Ca{\sc ii} H  & 4044 & 0.028 & $a$? & 0.8$\pm$0.2 & $-$1.9$\pm$0.6 \\
                    &       & Ca{\sc ii} K  & 4074 & 0.027 & $a$? & 0.8$\pm$0.2 & $-$2.0$\pm$0.6 \\

 & & & & & & & \\

1FGL J1926.8+6153   & 0.012 &               &      &       &      &             &                \\
                    &       & Ca{\sc ii} H  & 3979 & 0.011 & $a$? & 1.4$\pm$0.2 & $-$13$\pm$2    \\
                    &       & Ca{\sc ii} K  & 4015 & 0.012 & $a$? & 1.1$\pm$0.2 & $-$9.7$\pm$1.9 \\

 & & & & & & & \\

1FGL J1933.3+0723   & 0.010 &               &      &       &      &               &                \\
                    &       & Ca{\sc ii} H  & 3976 & 0.011 & $a$? & 1.3$\pm$0.2   & $-$3.2$\pm$0.5 \\
                    &       & Ca{\sc ii} K  & 4005 & 0.009 & $a$? & 0.84$\pm$0.17 & $-$2.1$\pm$0.4 \\

 & & & & & & & \\

1FGL J2042.2+2427   & 0.105 &               &      &       &      &                &               \\
                    &       & Ca{\sc ii} H  & 4348 & 0.105 & $g$  &    7.6$\pm$0.8 & $-$28$\pm$3   \\
                    &       & Ca{\sc ii} K  & 4381 & 0.104 & $g$  &    7.0$\pm$0.7 & $-$26$\pm$3   \\
                    &       & G-band        & 4749 & 0.103 & $g$  &    5.1$\pm$0.5 & $-$22$\pm$2   \\
                    &       & Mg b          & 5720 & 0.105 & $g$  &    7.4$\pm$0.7 & $-$39$\pm$4   \\
                    &       & Na D          & 6512 & 0.106 & $g$  &    4.4$\pm$0.4 & $-$24$\pm$2   \\
                    &       & [N {\sc ii}]  & 7276 & 0.105 & $e$  & $-$1.9$\pm$0.4 &   9.4$\pm$1.9 \\

\noalign{\smallskip} 
\hline
\noalign{\smallskip} 

\multicolumn{8}{l}{$^{\rm *}$: Type of spectral line ($e$: emission line; 
$g$: absorption line from the host galaxy;} \\
\multicolumn{8}{l}{$a$: absorption line from intervening system)} \\
\multicolumn{8}{l}{Note: fluxes are in units of 10$^{-16}$ erg cm$^{-2}$ 
s$^{-1}$ and are not corrected for intervening Galactic absorption.} \\

\noalign{\smallskip}
\hline
\hline
\noalign{\smallskip}
\end{tabular}
\end{center}
\end{table*}

\section{Discussion}

Stephen et al. (2010) suggested that the {\it Fermi}-{\it ROSAT} sample of 
possible BL Lac objects they selected may be made of objects with a very 
energetic electron distribution, such that the synchrotron emission peak 
of their spectral energy distribution (SED) falls in the ultraviolet band 
or in the soft X--rays. Thus, they may be high-synchrotron-peaked BL Lac 
(HBL) objects, and they could in principle be detectable up to the TeV 
bands (Padovani \& Giommi 1995; Fossati et al. 1998). Moreover, Shaw et 
al. (2013) classify as HBL the eight sources in common with our sample, 
again on the basis of their broadband SEDs. This inference is also 
supported by the relatively hard $\gamma$--ray photon spectral index shown 
by most BL Lac objects in the present sample ($\Gamma \la$ 2; see Table 2 
of Stephen et al. 2010).

In order to check and verify this assertion, we applied the method 
described in Abdo et al. (2010c,d) for the SED classification and the 
determination of the frequency of the synchrotron peak for all the BL Lacs 
in our sample. This can be made by computing the broadband spectral 
indices $\alpha_{\rm ro}$ and $\alpha_{\rm ox}$ between the radio and 
optical (5 GHz and 5000 \AA) and optical and X--ray (5000 \AA~and 1 keV) 
bands, respectively.

To this aim, we used the {\it ROSAT} X--ray fluxes reported in Stephen et 
al. (2010); we corrected them for the Galactic absorption according to the 
maps of Kalberla et al. (2005); and we assumed an X--ray power law 
spectral shape with photon index $\Gamma$ = 2.23 (Brinkmann et al. 1997). 
For the optical, we used the $R$-band magnitudes in Table 2; we corrected 
them for the intervening Galactic absorption using the maps of Schlegel et 
al. (1998) and the reddening law of Cardelli et al. (1989), together with 
the total-to-selective extinction ratio of Rieke \& Lebofsky (1985); we 
then determined the corresponding optical fluxes using the conversion 
factor of Fukugita et al. (1995) and we rescaled the values at 5000 
\AA~assuming a power law spectral shape $F(\lambda) = \lambda^\alpha$ with 
$\alpha$ = $-$0.7 (Falomo et al. 1993) for the optical continuum. We note 
that the choice of this average value rather than those given in Table 2 
does not substantially alter our results; one can see that a variation of 
$\pm$0.3 in $\alpha$ corresponds to a change of less than 2\% in both 
$\alpha_{\rm ro}$ and $\alpha_{\rm ox}$. Finally, we used the radio flux 
densities compiled in Stephen et al. (2010) and assumed a flat radio 
spectral shape (e.g., Begelman et al. 1984) to rescale them at 5 GHz.

The results for our sample of BL Lacs are graphically shown in Fig. 7, 
modeled following Fig. 4 of Abdo et al. (2010c): all objects fall below 
the thick broken line. This indicates that the frequency of their 
synchrotron peak, computed following Sect. 3.3.2 of Abdo et al. (2010c) 
and shown in Table 2, is above 10$^{14}$ Hz in all cases, thus they are 
at least intermediate-synchrotron-peaked BL Lac (IBL) objects. Moreover, 
18 of them have synchrotron peak frequency larger than 10$^{15}$ Hz (for 
1FGL J0848.6+0504 this value is $>$10$^{16}$ Hz) and thus can be 
classified as HBLs. One of the sources in our sample, 1FGL J0648.8+1516, 
has been detected up to the TeV range (Aliu et al. 2011); it is also among 
the ones with the largest synchrotron peak frequency within those studied 
in this paper (see the last column of Table 2).

This supports and extends the suggestion of Stephen et al. (2010) that 
this selection obtained by cross-correlating the {\it Fermi} and the {\it 
ROSAT} catalogs allowed pinpointing preferentially high-energy BL Lacs. 
Also, this partially agrees with what Shaw et al. (2013) find for the 
objects in common with our sample, because we classify 1FGL 
J1553.5$-$3116, 1FGL J1643.5$-$0646, and 1FGL J2146.6$-$1345 as IBLs 
rather than HBLs. However, given the approximations we introduced in our 
computations, we can nevertheless conclude that our results are compatible 
with those of Shaw et al. (2013).

It should also be noted that the recent catalog of {\it Fermi}/LAT sources 
detected above 10 GeV (1FHL; Ackermann et al. 2013) contains 20 out of the 
25 sources listed in Table 2 (the exceptions are 1FGL J0223.0$-$1118, 1FGL 
J0848.6+0504, 1FGL J1643.5$-$0646, 1FGL J1933.3+0723, and 1FGL 
J2323.0$-$4919). Eighteen of them are indeed associated by Ackermann et 
al. (2013; their Table 3) with the corresponding BL Lac objects considered 
in the present paper, while just two (1FGL J0841.4$-$3558 and 1FGL 
J1353.6$-$6640) appear unassociated with any longer-wavelength source. 
This connection can be considered an additional clue about the possibility 
of detecting TeV emission from the majority of these BL Lacs.

\begin{figure}[th!]
\mbox{\psfig{file=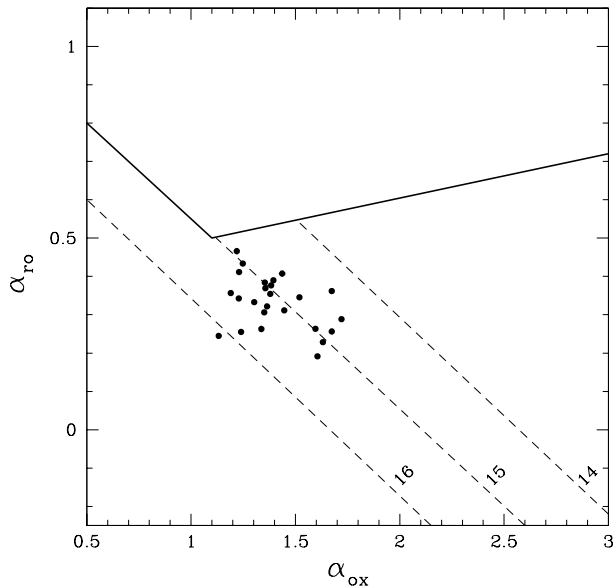,width=9cm,angle=0}}

\caption{Positions of the BL Lacs listed in Table 2 on the 
$\alpha_{\rm ox}$--$\alpha_{\rm ro}$ plane. The thick solid line 
indicates the region described in Sect. 3.3.2 of Abdo et al. (2010c). The 
dashed lines correspond to the loci in which the synchrotron peak 
frequency is 10$^{14}$, 10$^{15}$, and 10$^{16}$ Hz.}
\end{figure}

A further indication of the possible TeV nature of the BL Lac objects in 
the sample of Stephen et al. (2010) comes from the treatment of Massaro et 
al. (2013b). These authors, by means of their method (D'Abrusco et al. 
2013; Massaro et al. 2012, 2013a) of selecting BL Lac candidates among the 
unidentified {\it Fermi} sources using the {\it WISE} mid-infrared (MIR) 
magnitudes (Wright et al. 2010) of putative candidates, identified the MIR 
characteristics of a sample of possible TeV BL Lac objects. Massaro et al. 
(2013b) thereby determined a locus in the so-called blazar strip in which 
these possible TeV-emitting BL Lac objects are found in the {\it WISE} 
[4.6]-[12]/[3.4]-[4.6] color-color diagram (see Fig. 8).

When we consider the {\it WISE} colors of the objects in the sample of 
Stephen et al. (2010), we see that all objects we identify here as BL Lacs 
fall along the blazar strip and, most importantly, nearly all of them
are within the limits of this locus of TeV-emitting BL Lacs.
The only exception is 1FGL J2042.2+2427, which nevertheless lies close to 
the blue end of the strip (in the lower-left part of Fig. 8), thus 
perhaps indicating a prolongation of it to bluer colors and possibly an 
even more extreme BL Lac nature for this source.

\begin{figure}[th!]
\mbox{\psfig{file=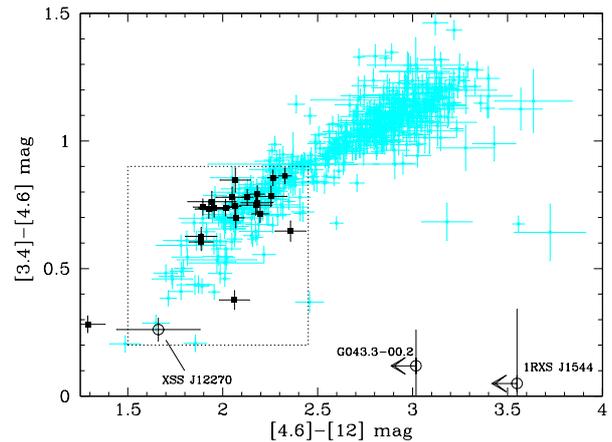,width=9cm,angle=270}}

\caption{The [4.6]-[12]/[3.4]-[4.6] MIR color-color plot reporting the 
positions of $\gamma$--ray emitting blazars (in cyan) associated 
with {\it WISE} sources forming the blazar strip (see Massaro et al. 
2013a for more details), together with the BL Lac objects (filled squares) 
identified in this paper and belonging to the {\it Fermi}-{\it ROSAT} 
sample of Stephen et al. (2010). In the diagram, the positions of 
non-blazar sources in the list of Stephen et al. (2010) are labeled and 
indicated with open circles. The dashed box indicates the locus of the
{\it WISE} blazar strip considered by Massaro et al. (2013b) as 
containing likely TeV-emitting BL Lacs.}
\end{figure}

We would like to point out here some necessary caveats concerning the 
application of this X--ray/GeV selection method. Although 25 out of 30 
sources of our {\it Fermi}/{\it ROSAT} sample are confirmed BL Lac sources 
(see Table 2), we see from the Appendix of this paper and from Stephen et 
al. (2010) that four objects are within the Galaxy, namely one SNR 
(G043.3$-$00.2), one LMXB (XSS J12270$-$4859), and two CVs (1RXS 
J083842.1$-$282723 and 1RXS J154439.4$-$112820, putatively associated with 
1FGL J0838.6$-$2828 and 1FGL J1544.5$-$1127, respectively), while one 
source (1FGL J2056.7+4938) is still unidentified. We can thus state that 
the {\it a posteriori} probability of success in pinpointing BL Lac-type 
objects using the approach of Stephen et al. (2010) is on the order of 
80\%. We remark that none of these five sources except 1FGL J1910.9+0906 
is present in the 1FHL catalog of Ackermann et al. (2013).

We also stress that the two CVs detected here are unlikely to be 
physically associated with the {\it Fermi} sources 1FGL J0838.6$-$2828 and 
1FGL J1544.5$-$1127; while it is known that accreting white dwarfs may 
produce outbursts of $\gamma$--rays (see Cheung 2013 for a review and Hays 
et al. 2013 for the most recent case), these are associated with powerful 
transient phenomena such as nova eruptions rather than with persistently 
emitting X--ray CVs. Possible TeV emission from some magnetic CVs was 
announced in the past (Bhat et al. 1991; Bowden et al. 1992; Meintjes et 
al. 1992), but was never confirmed by subsequent deeper observations (Lang 
et al. 1998; Sidro et al. 2008; L\'opez-Coto et al. 2013). Indeed, models 
describing production of $\gamma$--ray emission from magnetic CVs have 
been developed (e.g., Schlegel et al. 1995 and references therein; 
Bednarek \& Pabich 2011), but they can hardly be applied to the two cases 
in our sample as they do not appear to host a white dwarf with a strong 
magnetic field (see Appendix).

To corroborate on a statistical basis the lack of physical connection 
between the {\it ROSAT} and {\it Fermi} emissions in the case of the two 
CVs reported in the Appendix, we can consider the spatial density of CVs 
in the Galaxy from Rogel et al. (2008). We find that the number of CVs 
lying within 760 pc of the Earth (that is, the inferred distance of the 
farthest of the two CVs identified here; see Table A.1) and falling in a 
sky area of size equal to the sum of the error boxes of the 30 {\it 
Fermi}/LAT sources of Stephen et al. (2010; we conservatively assumed an 
average error circle of radius 6 arcmin) is about one. Given the 
uncertainties associated with the local Galactic CV density and the 
distance of the CVs reported in the Appendix, this value is fully 
compatible with our observational result, according to which we find two 
CVs among the sample considered in this paper. Because of all of the 
above, we deem these two putative associations in the list of Stephen et 
al. (2010) as spurious.

A possible way to identify ``BL Lac impostors" before performing optical 
spectroscopy on a X--ray/GeV selected sample such as the present one is 
the combined use of {\it WISE} and radio data. Indeed, as one can see from 
Fig. 8, {\it WISE} observations indicate that all BL Lacs in our sample 
fall along the blazar strip of Massaro et al. (2012, 2013a), with 1RXS 
J154439.4$-$112820 and 1FGL J1910.9+0906 lying well outside of it; 1RXS 
J083842.1$-$282723 and 1FGL J2056.7+4938 are not detected with {\it WISE}, 
the latter possibly because of source confusion introduced by the nearby 
bright star BD+49 3420 (see Sect. 2). The only apparent outlier is 1FGL 
J1227.9$-$4852, which falls both along the blazar strip and in the TeV box 
shown in Fig. 8; however, it can be discarded {\it a priori} as a blazar 
candidate because it was not detected in any of the main radio surveys 
(although deep radio observations revealed a faint emission from this 
source; see Hill et al. 2013). Thus, although the {\it WISE} TeV box could 
contain a small number of false positives, the additional use of cataloged 
radio information can help us to remove them.

\section{Conclusions}

The spectroscopic investigation of the putative optical counterparts of a 
sample of positionally correlated X--ray and $\gamma$--ray sources 
extracted from {\it ROSAT} and {\it Fermi} catalogs, respectively 
(Stephen et al. 2010), shows that 25 out of 30 are BL Lac-type AGNs. For 
eight of these objects we provide a redshift measurement, while for 
the remaining ones a lower limit to this observable is given. Two more 
sources are instead classified as Galactic CVs, and thus are quite likely 
spuriously associated with the GeV emission detected with {\it Fermi}. 
This suggests an 80\% {\it a posteriori} probability of finding BL Lac 
objects by cross-correlating X--ray and $\gamma$--ray catalogs.

Our analysis also hints to the fact that the synchrotron peak of the SED 
of BL Lac objects selected with this approach falls in the higher-energy 
side of the spectral range observed for this type of source (i.e., 
between the far ultraviolet and the soft X--rays) which suggests that this 
kind of selection may unearth objects detectable at TeV energies with 
instruments foreseeable in the near future (e.g., the \v{C}erenkov 
Telescope Array; Actis et al. 2011).

We conclude by pointing out that the interest in these extreme TeV blazars 
is driven by the possibility of obtaining information on both the 
acceleration processes of charged particles in relativistic flows (e.g., 
Ghisellini et al. 2010) and the intensity of the extragalactic background 
light, which reflects the time-integrated history of light production and 
re-processing in the universe (e.g., Dom\'{i}nguez et al. 2013); hence, 
its measurement can provide information on the history of cosmological 
star formation (Mankuzhiyil et al. 2010). This set of likely high-energy 
BL Lacs can thus be an initial sample to test the points mentioned above.

\begin{acknowledgements}

We thank Francesca Ghinassi and Avet Harutyunyan for Service Mode 
observations at the TNG; Duncan Castex and Ariel S\'anchez for assistance 
at the ESO NTT; Manuel Hern\'andez for Service Mode observations at the 
CTIO telescope and Fred Walter for coordinating them. NM thanks Ivo 
Saviane, Valentin Ivanov and Roberto Soria for help with data acquisition, 
Paola Grandi for useful discussions, and Francesco Massaro for help with 
the preparation of Figure 8.
We also thank the anonymous referee for useful remarks which helped us
to improve the quality of this paper. 
This research has made use of the ASI Science Data Center Multimission 
Archive and of data obtained from the ESO Science Archive Facility;
it also used the NASA Astrophysics Data System Abstract Service, 
the NASA/IPAC Extragalactic Database (NED), and the NASA/IPAC Infrared 
Science Archive, which are operated by the Jet Propulsion Laboratory, 
California Institute of Technology, under contract with the National 
Aeronautics and Space Administration.
This publication made use of data products from the Two Micron All 
Sky Survey (2MASS), which is a joint project of the University of 
Massachusetts and the Infrared Processing and Analysis Center/California 
Institute of Technology, funded by the National Aeronautics and Space 
Administration and the National Science Foundation.
This research has also made use of the SIMBAD and VIZIER databases operated 
at CDS, Strasbourg, France.
NM thanks the Departamento de Astronom\'{i}a y Astrof\'{i}sica of the
Pontificia Universidad Cat\'olica de Chile in Santiago for the warm
hospitality during the preparation of this paper.
PP and RL are supported by the ASI-INAF agreement No. I/033/10/0.
DM is supported by the BASAL CATA PFB-06 and FONDECYT No. 1130196 grants.
GG is supported by Fondecyt grant No. 1120195.
\end{acknowledgements}

\appendix

\section{CVs detected in the sample}

\begin{figure*}
\mbox{\psfig{file=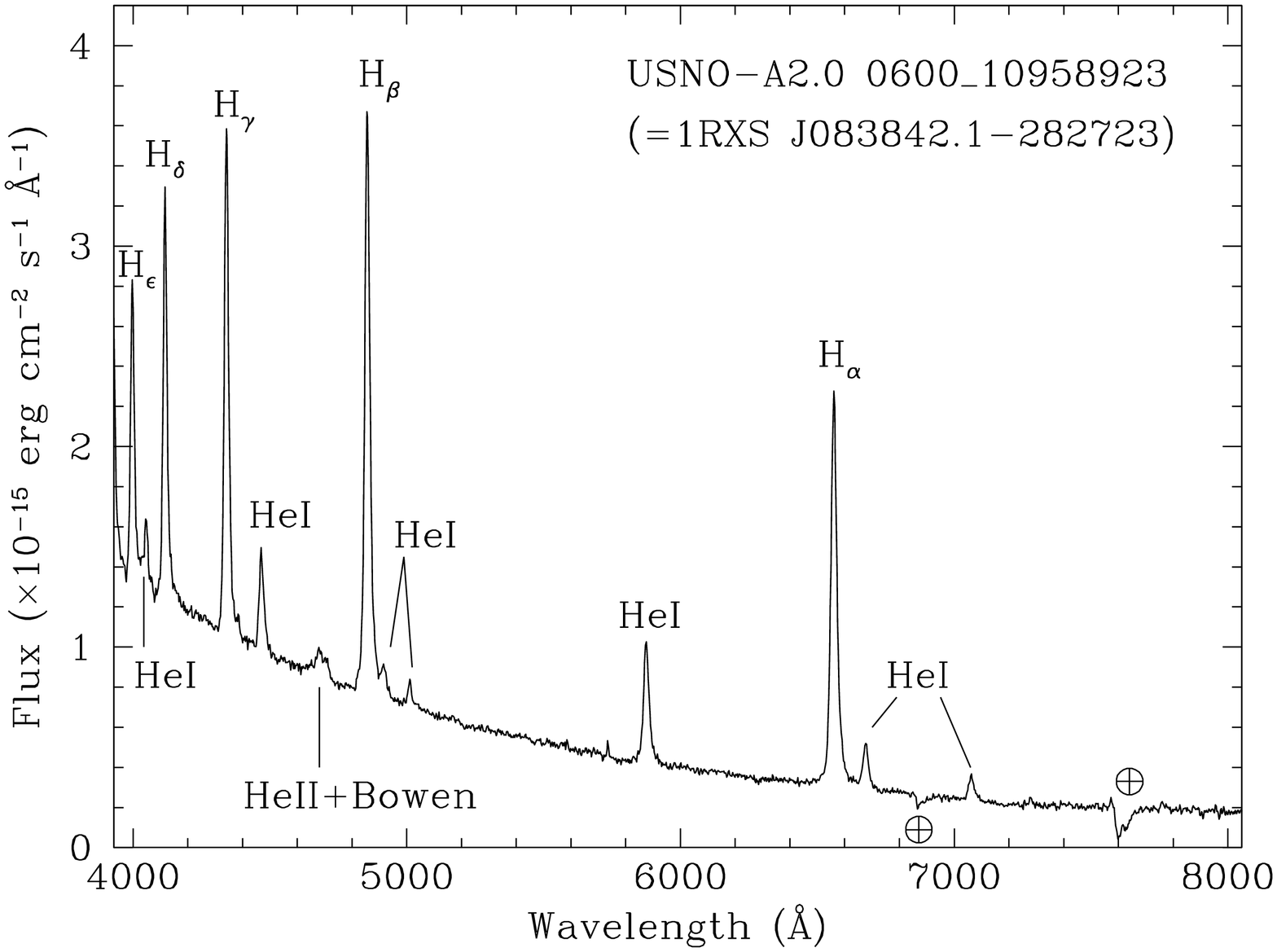,width=9cm,angle=270}}
\mbox{\psfig{file=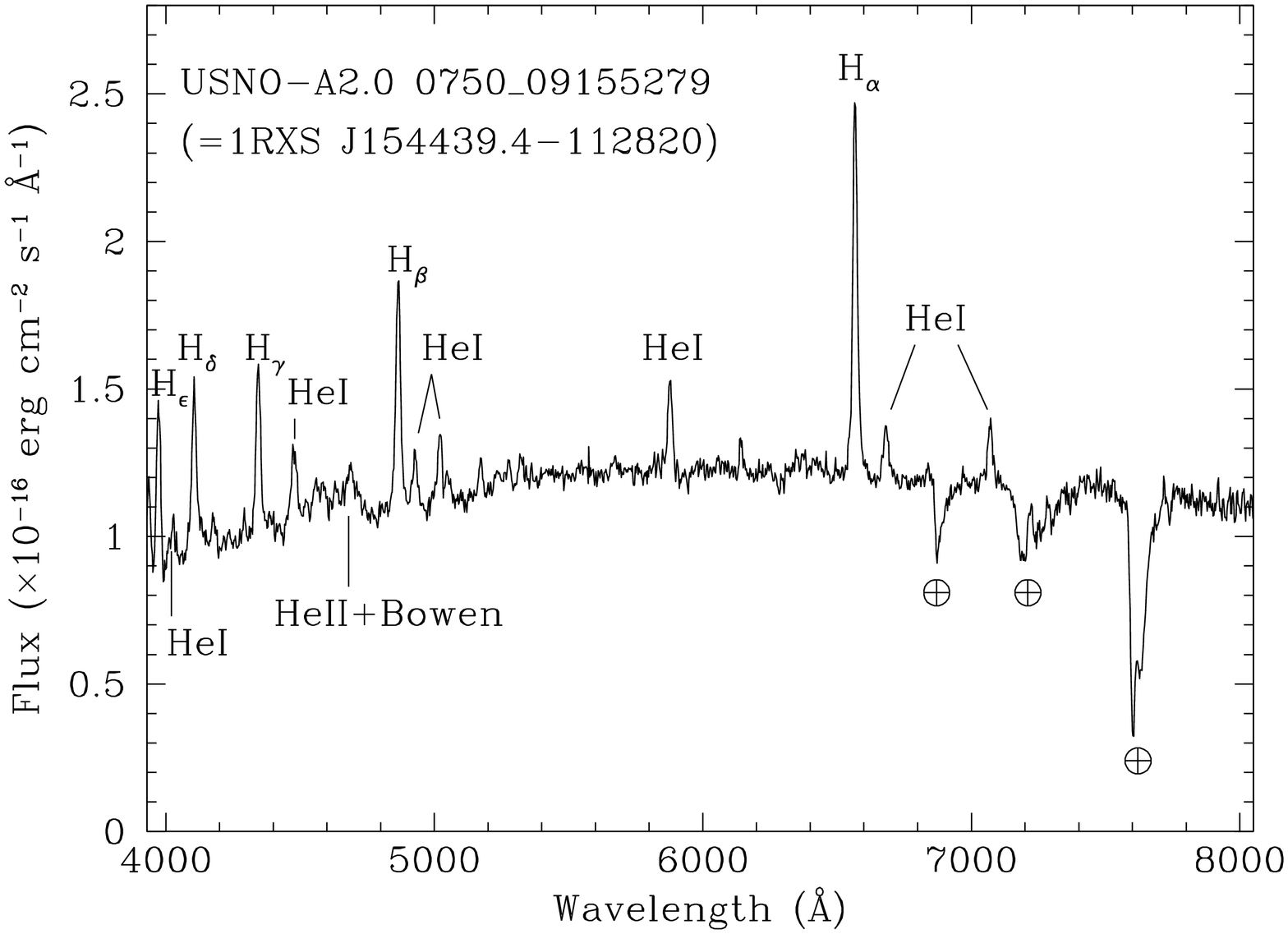,width=9cm,angle=270}}

\caption{Spectra (not corrected for the intervening Galactic absorption) 
of the optical counterparts of the two CVs identified within the sample of 
{\it ROSAT} sources considered in this paper. For each spectrum, the main 
spectral features are labeled. The symbol $\oplus$ indicates atmospheric 
telluric absorption bands.}
\end{figure*}

\begin{table*}
\caption[]{Synoptic table containing the main results concerning the 
two CVs identified in the present sample of {\it ROSAT} sources.}
\scriptsize
\vspace{-.3cm}
\begin{center}
\begin{tabular}{lccccccccr}
\noalign{\smallskip}
\hline
\hline
\noalign{\smallskip}
\multicolumn{1}{c}{Object} & \multicolumn{2}{c}{H$_\alpha$} & 
\multicolumn{2}{c}{H$_\beta$} & \multicolumn{2}{c}{He {\sc ii} $\lambda$4686} & 
$R^{\rm a}$ & $d$ & \multicolumn{1}{c}{$L_{\rm X}$} \\
\cline{2-7}
\noalign{\smallskip} 
 & EW & Flux & EW & Flux & EW & Flux & mag & (pc) & \\

\noalign{\smallskip}
\hline
\noalign{\smallskip}

1RXS J083842.1$-$282723 & 180$\pm$5 & 53.6$\pm$1.6 & 88$\pm$3 & 68$\pm$2 & 1.1$\pm$0.2 & 0.89$\pm$0.18 & 
 17.6 & $\sim$530 & 2.6 \\

& & & & & & & & & \\ 

1RXS J154439.4$-$112820 & 23.0$\pm$0.7 & 2.76$\pm$0.08 & 18.0$\pm$0.9 & 1.9$\pm$0.1 & 8.9$\pm$0.9 & 0.95$\pm$0.10 & 
 18.4 & $\sim$760 & 7.2 \\

\noalign{\smallskip} 
\hline
\noalign{\smallskip} 
\multicolumn{10}{l}{Note: EWs are expressed in \AA; line fluxes are
in units of 10$^{-15}$ erg cm$^{-2}$ s$^{-1}$; whereas the 0.1--2.4 keV X--ray luminosities
are in units of} \\
\multicolumn{10}{l}{10$^{31}$ erg s$^{-1}$.} \\
\multicolumn{10}{l}{$^{\rm a}$: from the USNO-A2.0 catalog} \\
\noalign{\smallskip} 
\hline
\hline
\noalign{\smallskip} 
\end{tabular} 
\end{center}
\end{table*}

Spectroscopy of the putative optical counterparts of two objects (1FGL 
J0838.6$-$2828 and 1FGL J1544.5$-$1127) associated by Stephen et al. 
(2010) with {\it ROSAT} sources 1RXS J083842.1$-$282723 and 1RXS 
J154439.4$-$112820, respectively, yields a surprising result. As one can 
see in Fig. A.1, both objects show several Balmer lines (up to 
H$_\epsilon$ at least) and helium lines in emission. In all cases these 
spectral features are at redshift $z$ = 0, which of course means that 
these objects lie within the Galaxy. Their spectral appearance is indeed 
typical of Galactic CVs (e.g., Warner 1995).

The main spectroscopic results and the main astrophysical parameters which 
can be inferred from the available observational data are reported in 
Table A.1. The determination of these parameters has been performed as 
follows.

The estimate of the reddening along the line of sight was attempted by 
considering an intrinsic H$_\alpha$/H$_\beta$ line ratio of 2.86 
(Osterbrock 1989) and by computing the corresponding color excess from 
the comparison between the intrinsic line ratio and the measured ratio by 
applying the Galactic extinction law of Cardelli et al. (1989).
However, since both spectra show an inverted Balmer ratio (especially in the 
case of 1RXS J083842.1$-$282723), we considered $A_V \sim$ 0 in both cases.

To derive the distance of the two CVs we assumed an absolute magnitude 
M$_V \sim$ 9 and an intrinsic color index $(V-R)_0 \sim$ 0 mag (Warner 
1995). Although this method gives an order-of-magnitude value for the 
distance of Galactic CVs, our past experience (Masetti et al. 2013 and 
references therein) tells us that these estimates are in general correct 
to within 50\% of the refined value subsequently determined with more 
precise approaches.

In Table A.1 the 0.1--2.4 keV X--ray luminosities of the two objects were 
obtained using the {\it ROSAT} fluxes as reported in Stephen et al. (2010)
and the thereby inferred distances to the sources.

It can be noted from Table A.1 that both sources present an He {\sc ii} 
$\lambda$4686 / H$_\beta$ EW ratio much smaller than 0.5; moreover, the EW 
of the He {\sc ii} $\lambda$4686 line is smaller than 10 \AA~in both 
cases. This is an indication that the two CVs are hosting a non-magnetized 
white dwarf (see Warner 1995). Because of the nature of these objects, we 
consider their physical association with the two above mentioned {\it 
Fermi} sources unlikely (see main text for details).

\end{document}